%% file: main.tex
\input{Definitions}
\begin{document}

\title{Automotive Radar Performance in Environments with Multiple Interference Sources}

\author{Oren~Longman,
    Guy~Mardiks,
    Tomer~Maayan,
    and~Gaston~Solodky
    \\
    General Motors, Technical Center - Israel
    \\
    oren.longman@gmail.com
}

\maketitle

\begin{abstract}
    Automotive radars are increasingly susceptible to mutual interference from neighboring radar systems, which can lead to false target detections and the masking of valid targets. While current interference levels remain manageable due to the relatively low penetration of radar-equipped vehicles, this assumption is expected to break down as radar adoption and per-vehicle radar density continue to increase.
This paper presents a comprehensive analysis of automotive radar performance in high-density interference environments. A realistic end-to-end simulation framework is developed at the intermediate frequency (IF) level, incorporating analytical interference modeling and detailed radar signal processing. The study evaluates the impact of interference across a range of future scenarios characterized by increased radar density and multiple radar configurations per vehicle.
Conventional interference mitigation techniques are systematically assessed to validate the simulation results, controlled experiments were conducted using a host radar exposed to up to 30 interfering radars in both anechoic and real-world environments.
The results demonstrate significant performance degradation under high interference conditions, with substantial reductions in detection probability and effective range. Among the evaluated techniques, time-frequency coding consistently provides the most robust performance, maintaining high detection probability even at elevated radar penetration rates. These findings highlight the limitations of current mitigation approaches and emphasize the need for coordinated and scalable interference management strategies in future automotive radar systems.
\end{abstract}


\IEEEpeerreviewmaketitle

\let\thefootnote\relax\footnotetext{Authors’ current addresses: Oren~Longman,
	 Guy~Mardiks, Tomer~Maayan, and~Gaston~Solodky are with General Motors, Technical Center - Israel, Herzelia 4672513, Israel (e-mail: orenlongman@gmail.com)}.

\section{Introduction}
\IEEEPARstart{M}{odern} vehicles are increasingly equipped with advanced driver assistance systems (ADAS) and autonomous vehicle (AV) technologies, including adaptive cruise control (ACC), automatic emergency braking (AEB), pedestrian detection/avoidance, lane departure warning/correction, blind-spot detection, automatic parking, and autonomous valet parking \cite{bengler2014three}. AV capabilities are categorized into levels from 0 to 5 \cite{sae2014taxonomy}, reflecting increasing self-driving capability in various conditions with reduced human supervision. These functionalities rely heavily on a suite of complementary sensors, primarily cameras, lidar, ultrasonic sensors, and radar to enable robust perception, decision-making, and control.

Among these sensing modalities, radar plays a central role due to its ability to provide reliable range and velocity measurements under adverse environmental conditions such as rain, fog, and low illumination. In addition, radar systems offer a favorable trade-off between cost, performance, and form factor. However, despite these advantages, automotive radars suffer from inherent limitations, including relatively low angular resolution, susceptibility to multipath effects, and vulnerability to interference from other radar systems \cite{brooker2007mutual}. Cameras provide high angular resolution, and are cost effective, but lack range information and are sensitive to weather and lighting.
Lidar offers good angular resolution with range, though rotating lidars are costly with large form factor. Recent developments include solid-state lidars that are more affordable and have a smaller form factor, but with reduced performance \cite{zhao2019recent} and ultrasonic sensors are limited in resolution and range \cite{carullo2001ultrasonic}.

\begin{figure}
    \centering
    \includegraphics[width=85mm,trim={20mm 20mm 20mm 20mm},clip]{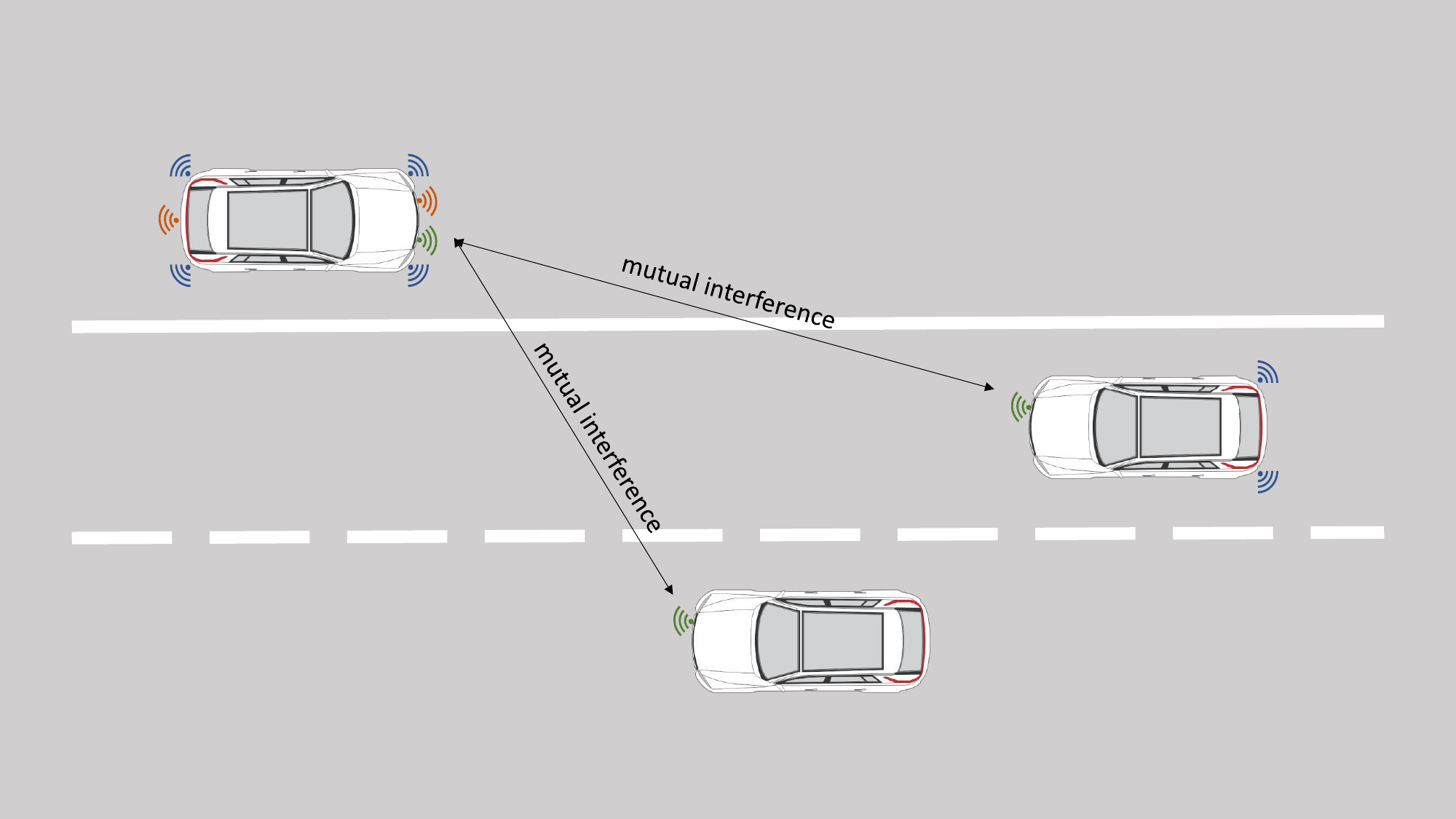}
    \caption{Vehicles with various radar installation typologies. The full topology consists of seven radars: a single front LRR (green), two front and back SRRs (red) and four  corner SBZAs (blue). The partial topology utilizes the front LRR and the two back SBZAs. The front topology utilizes only the front LRR.  The radar transmission interferes the other radars. \label{fig:interference_topology}}
    \captionsetup{justification=centering}
\end{figure}

The widespread adoption of radar-based sensing in modern vehicles is primarily driven by the implementation of ADAS features such as AEB, ACC, and lane-changing assistance \cite{bengler2014three}. These functionalities rely on different radar types, including long-range radar (LRR), short-range radar (SRR), and side blind zone alert (SBZA) radar, and topology suited to its specific requirements, as illustrated in Fig. \ref{fig:interference_topology}. The various radar types differ in terms of maximal detection range and vehicle mounting position, as well as in range resolution, maximal unambiguous Doppler and field of view (FOV). 

Regulatory requirements for equipping vehicles with AEB systems as a standard feature have already been implemented in regions including the United States, the European Union, and Japan \cite{NAAEB, EUAEB, JapanAEB}. ACC, utilizing a front radar, is becoming more common in new vehicles, while lane-changing assistance, which relies on SBZA radars, is also gaining popularity. AV technology further amplifies this trend, as multiple radars are often required for full functionality. This progression is transforming the automotive landscape, shifting from a minority of radar-equipped vehicles to a future where the majority will rely on radar systems.

In such environments, mutual interference between radar systems becomes a critical challenge. As an active sensing modality, radar systems continuously transmit electromagnetic signals, which may be received by neighboring radars operating in the same frequency bands, leading to interference \cite{kim2018peer,norouzian2021phenomenology}. These interfering signals can disrupt radar processing, leading to false alarms or preventing the detection of targets. While current systems often assume that interference events are relatively infrequent, this assumption is no longer valid in high-density scenarios, where both the likelihood and severity of interference increase significantly.

A range of interference mitigation strategies has been proposed to address this challenge, including avoidance-based approaches, filtering techniques, and structural or coordinated methods. Avoidance strategies aim to reduce interference through adaptive waveform selection, scheduling, or cognitive techniques \cite{hakobyan2019interference,  gurbuz2019overview, steck2018cognitive}. Structural strategies entail the harmonization and collaboration \cite{zhang2020vanet, wang2023performance,solodky2020cdma, deng2013interference}.

Several studies have focused on automotive radar interference analysis and mitigation techniques have provided valuable insights into interference characterization and mitigation \cite{aydogdu2020radar, ossowska2020imiko, toth2021analysis, alhumaidi2021interference,  uysal2018mitigation, liu2020decentralized,solodky2021clean}. The MOSARIM (MOre Safety for All by Radar Interference Mitigation) report and the Radar Congestion Study \cite{kunert2012eu, kunert2012d5, buller2018radar} have focused on interference analysis and mitigation but often rely on simulations or probabilistic methods rather than comprehensive end-to-end analysis \cite{munari2018stochastic}.

This paper presents a comprehensive evaluation of conventional interference mitigation techniques in future automotive environments characterized by increased radar density, both per vehicle and across the traffic ecosystem. The analysis considers diverse interference scenarios involving vehicles equipped with multiple radar types, including LRR, SRR, and SBZA. In particular, the scalability of mitigation strategies proposed in the MOSARIM report is systematically assessed as radar density increases.

To complement the simulation study, experimental validation was performed using three radar systems as hosts and an array of $30$ interfering radars, emulating the simulated conditions and enabling direct comparison between measured and simulated results. The experimental findings confirm the accuracy of the simulation framework and highlight the pronounced vulnerability of automotive radar systems to interference, underscoring the need for robust and scalable mitigation solutions.

The analysis focuses on linear frequency modulated continuous wave (LFM-CW) radar systems, while noting that other waveform types, such as phase-modulated continuous wave (PMCW), may exhibit similar interference effects with distinct characteristics \cite{yildirim2019impact,mazher2024automotive}.

Building on the above discussion, the main contributions of this work are summarized as follows:
\begin{enumerate}
    \item \textbf{Large-scale experimental validation:} 
    We present controlled interference experiments involving up to $30$ simultaneous interfering automotive radars in both anechoic chamber and real-world outdoor environments.

    \item \textbf{Simulation-to-measurement consistency analysis:} 
    An IF-level interference simulation framework is validated through direct comparison with field measurements under identical waveform configurations, demonstrating strong agreement (within $\sim1.5$dB).

    \item \textbf{Quantitative characterization of interference scaling:} 
    The impact of increasing numbers of interfering radars on noise floor, SNR, probability of detection, and maximum detection range is systematically analyzed across multiple radar systems.

    \item \textbf{Processing-chain-level interference analysis:} 
    Using raw ADC data, the propagation of interference through the radar processing chain (time domain, range FFT, and Doppler FFT) is analyzed.

    \item \textbf{Detector-dependent performance evaluation:} 
    The impact of interference on fixed-threshold and CA-CFAR detectors is experimentally compared.

    \item \textbf{Impact of waveform similarity:} 
    The results demonstrate that waveform similarity between radars leads to more severe interference effects.
\end{enumerate}

The remainder of this paper is organized as follows. Section \ref{sec:problem} describes the interference phenomenology and underlying signal model. Section \ref{sec:simulation} details the simulation architecture and modeling approach. Section \ref{sec:analysis} presents the simulation results under various interference scenarios. Section \ref{sec:experiment} provides the experimental validation in both anechoic chamber and field environments. Finally, Section \ref{sec:con} concludes the paper.

\section{Interference phenomenology and Model}\label{sec:problem}
Radar operation involves the transmission of electromagnetic signals, the reception of echoes reflected from targets, and signal processing. In addition to target echoes, interference signals from other radars may also be received. The impact of such interference depends on the characteristics of the interfering signal, the host radar's analog receiver, and its signal processing chain \cite{goppelt2010automotive,kim2018peer,fischer2013detection,norouzian2021phenomenology}. 

A single chirp of the host LFM-CW radar is defined as:
\begin{equation}
v(t) = 
\begin{cases}
    e^{-j2\pi (f_vt+\frac{1}{2}\alpha_v {t}^2)}, & \text{if}\ 0 \leq t \leq T_v \\
    0, & \text{otherwise}
\end{cases} \;,
\end{equation}
where $f_v$ is the host carrier frequency, $\alpha_v$ is the chirp slope, and $T_v$ is the chirp duration. For simplicity of presentation, the chirp start time is assumed to be $0$; however, in the full model, the start time is randomized as described later.
A single chirp of an interference LFM-CW radar is defined as:
\begin{equation}\label{interference_chirp}
    I(t) =  V(t-\tau)
\end{equation}
where $\tau$ denotes the time offset between the host radar transmission and the reception of the interfering signal.

In frequency-modulated continuous wave (FMCW) systems, the received signal is mixed with the conjugate of the transmitted signal in a process known as stretch processing \cite{bilik2019rise}. For the case in which the received signal corresponds to the interference chirp in \eqref{interference_chirp}, the mixed signal is given by:

\begin{equation}
    s(t) = I(t) \cdot {v(t)}^* =
    \begin{cases}
        x(t), & \text{if}\ 0 \leq t \leq T_m \\
        0, & \text{otherwise}
    \end{cases} \;,
\end{equation}

where $T_m = \text{min}(T_v,T_i)$ and $x(t)$ is expressed as:
\begin{equation}
    x(t) = e^{-j2\pi (f_i(t-\tau)+\frac{1}{2}\alpha_i (t-\tau)^2)} e^{j2\pi (f_vt+\frac{1}{2}\alpha_v {t}^2)} \;,
\end{equation}

Neglecting constant phase terms yields:
\begin{equation} \label{radar_mixed}
    x(t) = \ e^{j2\pi (f_mt+\frac{1}{2}\alpha_m t^2)} \;,
\end{equation}

where $f_m = f_v-f_i+\alpha_i \tau$ and $\alpha_m = \alpha_v-\alpha_i$. 

\section{Simulation}\label{sec:simulation}
The impact of interference on radar performance is evaluated through simulation in an automotive environment where vehicles are equipped with radar systems. The scenario consists of a host vehicle with a radar-under-test and multiple interfering radars mounted on surrounding vehicles.

The simulation is numerical and models the host radar at the intermediate frequency (IF) level, following stretch processing \cite{bilik2019rise}. The radar system is modeled end-to-end, from signal transmission and reception to target detection, incorporating the antenna, RF receiver, sampling stage, and signal processing chain. The transmission mechanisms of the interfering radars are also modeled, with both host and interfering radars subject to clock drift \cite{rao2020interference}. Signal propagation is influenced by vehicle positions, which may block propagation paths, as well as by highway walls that introduce reflections.

Highway scenarios are generated based on the 3GPP model \cite{3gpp2019}, featuring a six-lane road configuration with three lanes in each direction.

Further details on each simulation component are provided in the following subsections.

\subsection{Simulation Scenario}
The simulation is based on a highway environment featuring six lanes (three in each direction) bordered by concrete walls. Vehicles are randomly distributed along the roadway according to the 3GPP model \cite{3gpp2019}. An alternative vehicle placement approach is based on a Poisson point process model \cite{al2017stochastic}. Vehicles are categorized as cars or trucks with dimensions of $2\text{m} \times 5\text{m}$ and $2.6\text{m} \times 13,\text{m}$ (width $\times$ length), respectively.

\begin{figure}
    \centering
    \begin{subfigure}{0.24\textwidth}
        \centering\captionsetup{width=.9\linewidth}%
        \begin{tikzpicture}
            \node[anchor=south] (img) {\includegraphics[width=\textwidth]{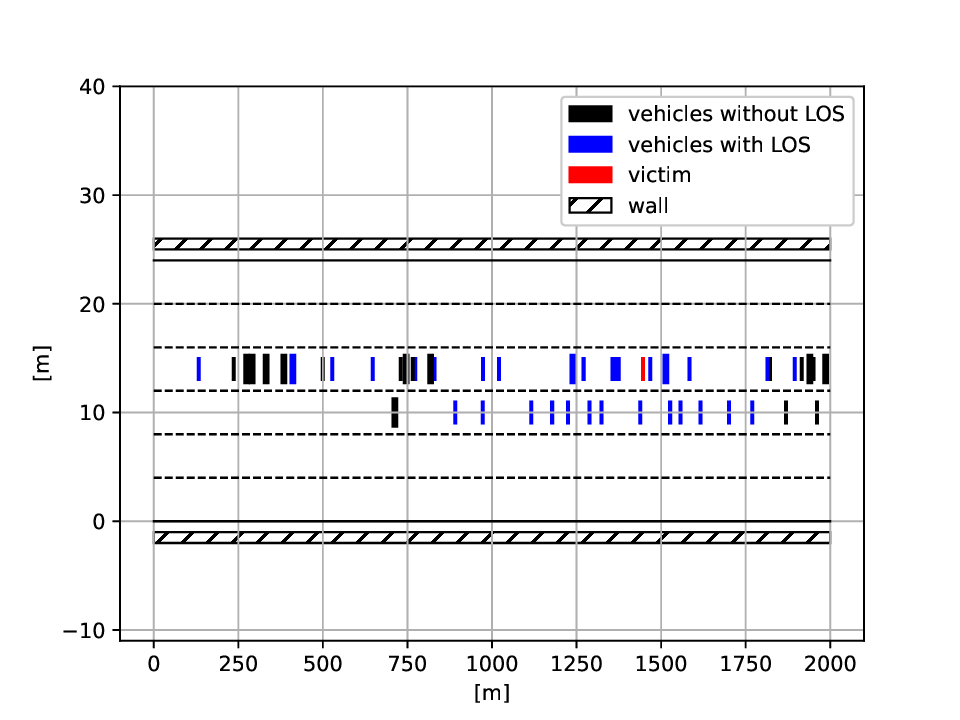}};
            \draw[black, thick] (-1.48,1.615) -- (1.59,1.615);
        \end{tikzpicture}
        \caption{Low density scenario where only 31 out of 49 vehicles have LOS.}
        \label{fig:scenario_low_LOS}
    \end{subfigure}
    \begin{subfigure}{0.24\textwidth}
        \centering\captionsetup{width=.9\linewidth}%
        \begin{tikzpicture}
            \node[anchor=south] (img) {\includegraphics[width=\textwidth]{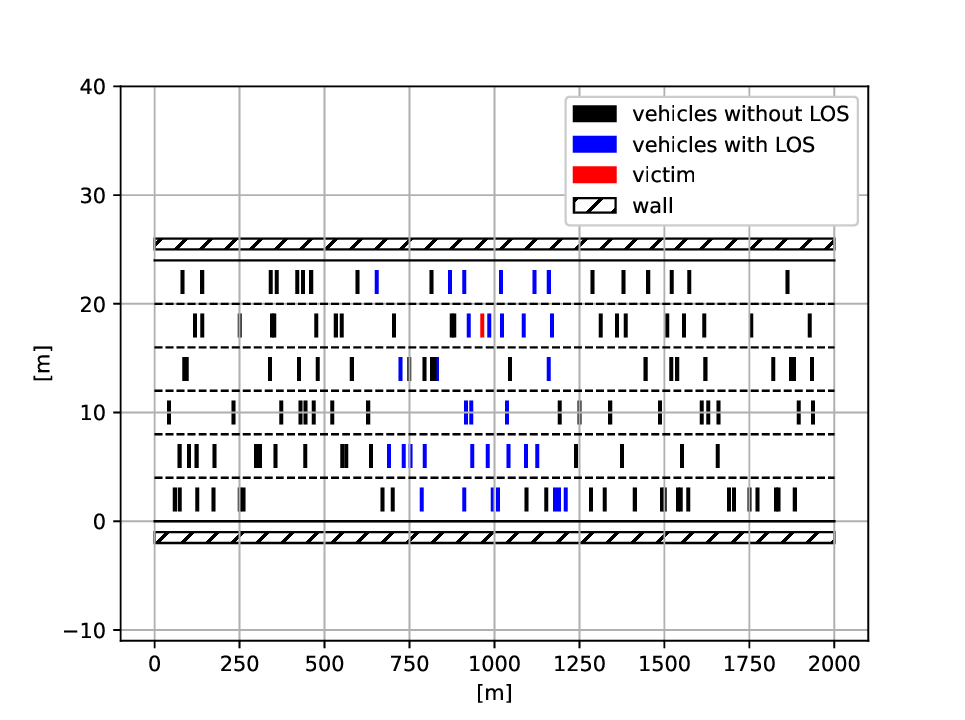}};
            \draw[black, thick] (-1.48,1.615) -- (1.59,1.615);
        \end{tikzpicture}
        \caption{Medium density scenario where only 33 out of 143 vehicles have LOS.}
        \label{fig:scenario_mid_LOS}
    \end{subfigure}
    \begin{subfigure}{0.24\textwidth}
        \centering\captionsetup{width=.9\linewidth}%
        \begin{tikzpicture}
            \node[anchor=south] (img) {\includegraphics[width=\textwidth]{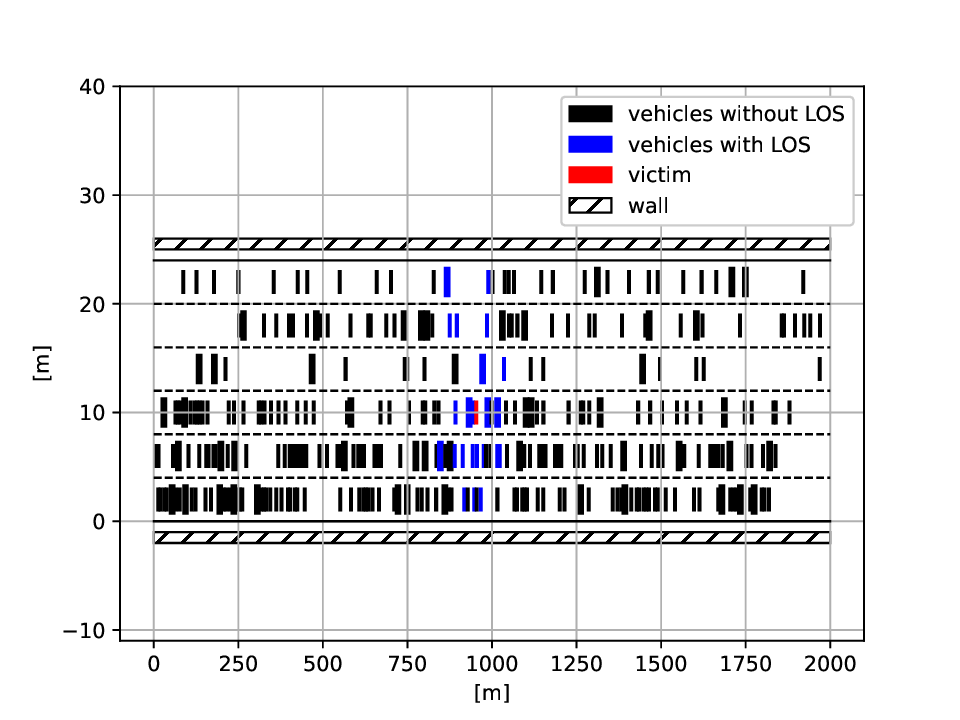}};
            \draw[black, thick] (-1.48,1.615) -- (1.59,1.615);
        \end{tikzpicture}
        \caption{High density scenario where only 23 out of 334 vehicles have LOS.}                   
        \label{fig:scenario_high_LOS}
    \end{subfigure}
    \caption{Illustrated scenarios with vehicle placement according to the 3GPP model. The host vehicle is shown in red, vehicles with LOS are shown in blue, and vehicles without LOS are shown in black.}
    \label{fig:scenarios_with_LOS}
\end{figure}
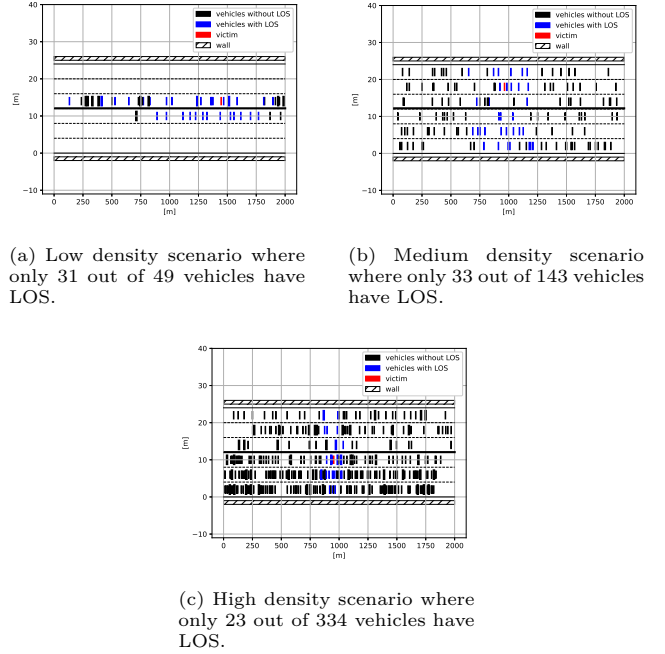

The study considers three scenarios with different vehicle densities: $49$, $143$, and $334$ vehicles, as illustrated in Fig. \ref{fig:scenarios_with_LOS}, sub-figures (a), (b) and (c), respectively. In the figures, blue rectangles represent interfering vehicles with line of sight (LOS), black rectangles denote interfering vehicles without LOS
and the red rectangle indicates the host vehicle.

\begin{figure}
    \centering
    \begin{tikzpicture}
            \node[anchor=south] (img) {\includegraphics[width=75mm,trim={2mm 2mm 2mm 2mm},clip]{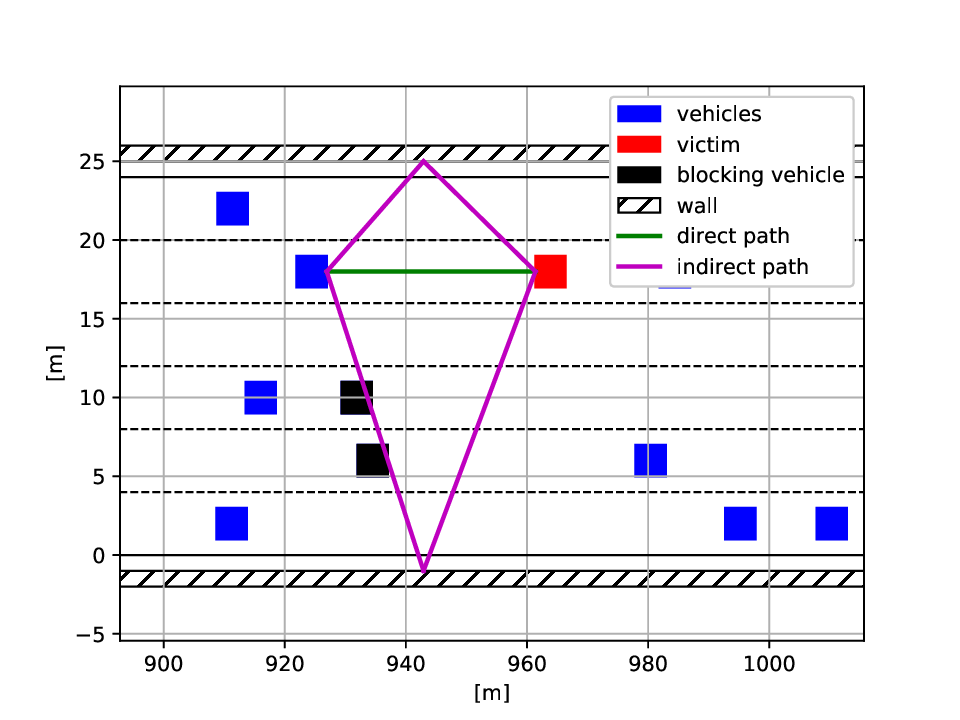}};
            \draw[black, thick] (-2.88,2.86) -- (3.07,2.86);
    \end{tikzpicture}
    \caption{Scenario illustrating road-wall reflections and vehicle blockage.}
    \label{fig:scenario_mp}
    \captionsetup{justification=centering}
\end{figure}

\begin{figure}
    \centering
    \includegraphics[width=90mm,trim={2mm 2mm 2mm 2mm},clip]{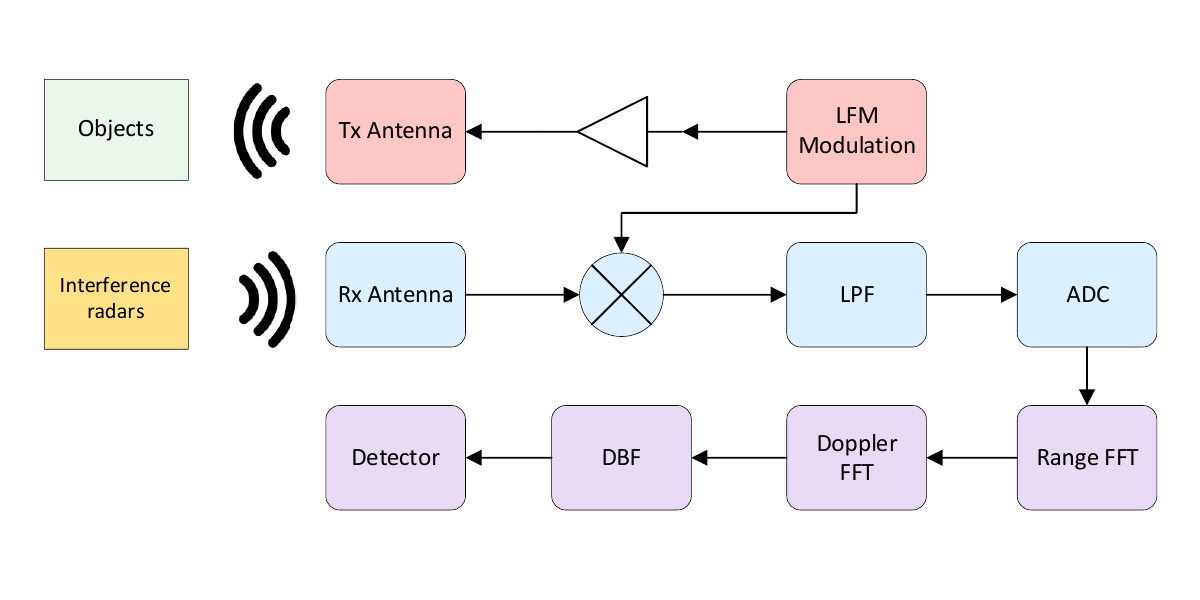}
    \caption{Radar system process flow.}
    \label{fig:radar_system}
    \captionsetup{justification=centering}
\end{figure}

\subsection{Propagation Model}
Obstacles between the host radar and interfering radars significantly affect signal propagation. Free-space propagation occurs between unobstructed objects \cite{boban2019multi, kamann2018automotive}. In automotive environments, various objects, including vehicles, pedestrians, bicycles, motorcycles, trucks, and infrastructure, can reflect, scatter, or block the signal \cite{kurz2021road, chipengo2018full}. In this simulation, two obstacle types are considered: road walls, which introduce reflections \cite{schipper2015simulative, terbas2019radar}, and vehicles, which act as blocking objects \cite{kui2021interference}.

Direct and indirect propagation paths are excluded if obstructed by vehicles, as illustrated in Fig. \ref{fig:scenario_mp}. The figure shows a direct path (green) and two indirect paths (purple) reflected by the upper and lower walls. The lower indirect path is blocked by vehicles and is therefore disregarded.

Due to blockage effects, many of the vehicles in the scenarios do not have LOS to the host vehicle radar, as shown in Fig. \ref{fig:scenarios_with_LOS}. In the low density scenario, $31$ out of $49$ vehicles have LOS (Fig. \ref{fig:scenario_low_LOS}), in the medium density scenario, $33$ out of $143$ vehicles have LOS (Fig. \ref{fig:scenario_mid_LOS}), in the high density scenario, $23$ out of $334$ vehicles have LOS (Fig. \ref{fig:scenario_high_LOS}).

The road walls are modeled as concrete, with a refractive index of $n=2.52-j0.076$ for $76.7$GHz, corresponding to a $3$cm thick concrete plate. This value is derived from averaged measurements of the complex reflection coefficient at $57.5$GHz and $95.9$GHz \cite{sato1995measurements}. The Fresnel’s reflection coefficient is given by:
\begin{equation} \label{parallel}
    R'_p = \frac{n^2cos \theta - \sqrt{n^2-sin^2\theta}}{n^2cos \theta + \sqrt{n^2-sin^2\theta}} \;,
\end{equation}
where $R'_p$ represent reflection coefficient for vertically polarized radar signals at the air-concrete interface.
    
\subsection{Radar System}
The radar system transmits an LFM-CW signal and receives echoes from targets along with interference from other radars. The received signal is processed through the chain illustrated in Fig. \ref{fig:radar_system}, which includes range FFT, Doppler FFT, and a cell-averaging constant false alarm rate (CA-CFAR) detector.

This interference signal is passed through a low-pass filter (LPF), with a cutoff frequency at $f_s$. If the signal frequency band $[f_m,f_m+t\alpha_m]$ lies outside the LPF bandwidth, it does not affect the host radar \cite{rao2020interference}. Additionally, if the chirps of the interfering radar do not overlap with those of the host radar, no interference occurs.

The time offset $\tau$ between interference radar chirps and host radar chirps varies across dwells and affects the degree of overlap between chirps.

\section{Analysis} \label{sec:analysis}
The simulation evaluates the impact of interference on radar performance across a range of automotive scenarios, focusing on various mitigation techniques and system configurations. The performance of the host radar is quantified in terms of probability of detection (PD), evaluated both with and without interference mitigation. Key parameters include radar type, vehicle density, installation topology, and mitigation strategy.

\subsection{Methodology}
Three scenarios were taken into account, representing low, medium, and high density environments, as illustrated in Figs. \labelcref{fig:scenario_low_LOS,fig:scenario_mid_LOS,fig:scenario_high_LOS}, with $49$, $143$, and $334$ vehicles, respectively. Each scenario is simulated over a duration of 10 seconds.

Three radar types are analyzed: LRR, SRR, and SBZA. Their key parameters, including the pulse repetition interval (PRI), slope, $T_c$ (chirp duration),  $f_c$ (carrier frequency), \#Chirps (number of chirps per dwell), frames per second (FPS), \#Elements (number of transmit and receive elements), and $P_t$ (transmit element power), are uniformly randomized within the ranges specified in Table \ref{tab:waveforms}. An element may represent a single patch, a patch array, or another configuration, thereby affecting the overall antenna gain.

\begin{table}
    \centering
    \caption{Radar waveform parameters}
    \label{tab:waveforms}
    \begin{tabular}{|c|cc|cc|cc|c|}
        \hline
        & \multicolumn{2}{c|}{LRR} & \multicolumn{2}{c|}{SRR} & \multicolumn{2}{c|}{SBZA} & Units \\
        & min & max & min & max  & min & max & \\
        \hline
        PRI & 18 & 20 & 22 & 27 & 30 & 35 & $\mu s$  \\
        Slope & 9 & 11 & 27 & 33 & 35 & 39  & $MHz \slash \mu s$ \\
        $T_c$ & 14 & 16 & 15 & 20 & 20 & 25  & $\mu s$ \\
        $f_c$ & 77 & 81 & 77 & 81 & 77 & 81  & $GHz$ \\
        \#Chirps & 256 & 512 & 128 & 256 & 128 & 256 & \\
        FPS & 25 & 30 & 25 &30 & 25 & 30 & $Hz$ \\
        $\#$Elements & 12 & 12 & 8  & 8 & 4 & 4 &   \\
        $P_t$ & 10 & 10 & 10 & 10 & 10 & 10 & $dBm$ \\
        Gain & 14 & 14 & 12.8 & 12.8 & 12.8 & 12.8  & $dBi$ \\
        \hline
    \end{tabular}
\end{table}

Three vehicle installation configurations, corresponding to increasing levels of ADAS and AV functionality, are evaluated: front (single radar), partial (three radars), and full (seven radar), as illustrated in Fig. \ref{fig:interference_topology}. The front topology includes a single forward-facing LRR supporting features like ACC and AEB. The partial topology adds two rear-mounted SBZA radars angled at $45^\circ$, enabling lane-change assistance. The full topology includes seven radars, extending the partial topology with additional forward- and backward-facing SRR units and two additional SBZA radars at the front, also angled at $45^\circ$, enabling more advanced autonomous driving capabilities when combined with other sensors.

The analysis includes four interference mitigation techniques derived from the MOSARIM project \cite{kunert2012d5}are incorporated into the host radar processing chain:  1) The predefined frequency technique allocates distinct frequency bands based on radar placement, reducing interference at the cost of reduced available bandwidth. 2) The predefined polarization technique, which employs orthogonal antenna polarizations (e.g., vertical and horizontal), providing approximately $15$dB interference suppression, though its effectiveness is limited by polarization changes due to reflections. 3) Time dithering technique which introduces random time delays between chirps to reduce interference between similar waveforms. 4) The time-frequency coding technique establishes orthogonal time-frequency resources with synchronization (e.g., via GNSS), enabling improved spectrum sharing and interference resilience.

\subsection{Results}
The analysis results, presented in Fig. \ref{fig:analysis_pd} show the PD as a function of the radar penetration rate, defined as the proportion of vehicles equipped with radar systems. The PD is computed as the ratio of successful target detections to the total number of dwells.

Fig. \ref{fig:pd_conv} presents the baseline performance without interference mitigation. As anticipated, PD degrades significantly with increasing radar penetration rates, particularly in high-density scenarios. Counterintuitively, the medium-density scenario exhibits lower PD than the high-density scenario, primarily due to increased signal blockage in the latter. The full topology shows the steepest degradation, followed by partial and front topologies. While low-density scenarios maintain PD above 95\%, high penetration rates reduce PD to below 20\%, severely impacting ADAS and AV functionality. Implementing interference mitigation techniques can alleviate PD degradation.

\subsubsection{Predefined Frequency Technique}
Fig. \ref{fig:pd_predefined_freq} illustrates the PD performance with the predefined frequency method. This method improves interference mitigation compared to the baseline, however, its effectiveness diminishes at high penetration rates, particularly in medium- and high-density scenarios with full radar configurations for the SBZA host radar.

\subsubsection{Predefined Polarization Technique}
Fig. \ref{fig:pd_predefined_pol} shows that the predefined polarization technique offers moderate improvement over the baseline but remains less effective than the predefined frequency technique. Its performance is limited by reduced orthogonality and sensitivity to polarization changes induced by reflections.

\subsubsection{Time Dithering Technique}
Fig. \ref{fig:pd_time_dithering} demonstrates that time dithering method provides negligible improvement. The randomized waveform configurations used in the simulation reduce the likelihood of persistent interference patterns, thereby limiting the effectiveness of this technique.

\subsubsection{Time-Frequency Coding Technique}
\begin{figure}
    \centering
    \begin{subfigure}[t]{0.24\textwidth}
        \centering
        \includegraphics[width=\textwidth]{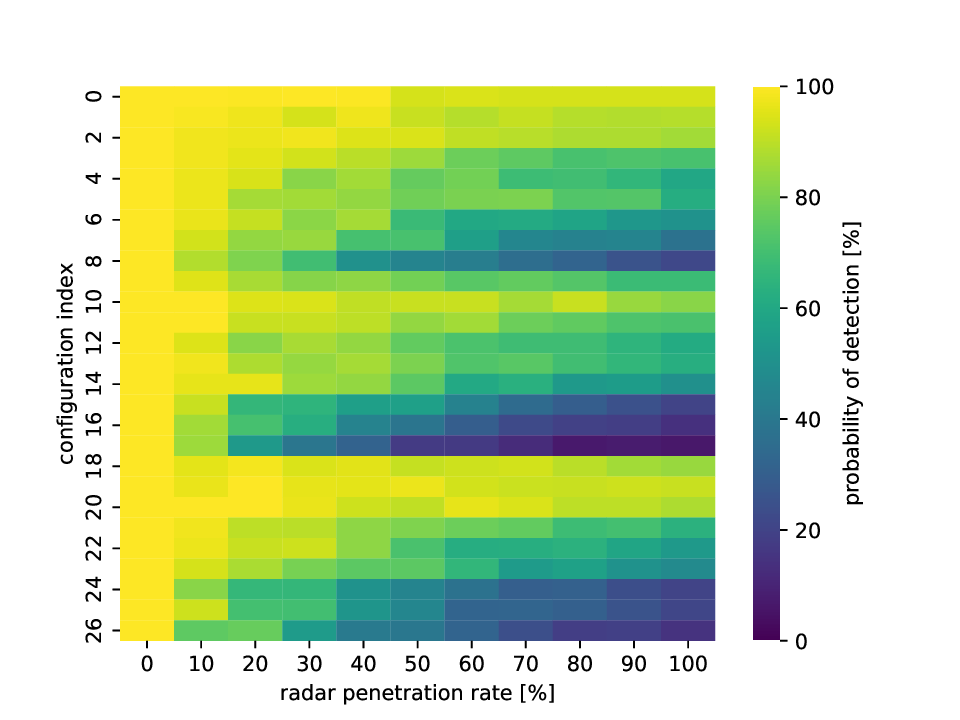}
        \caption{CA-CFAR method.}
        \label{fig:pd_conv}
    \end{subfigure}
    \begin{subfigure}[t]{0.24\textwidth}
        \centering 
        \includegraphics[width=\textwidth]{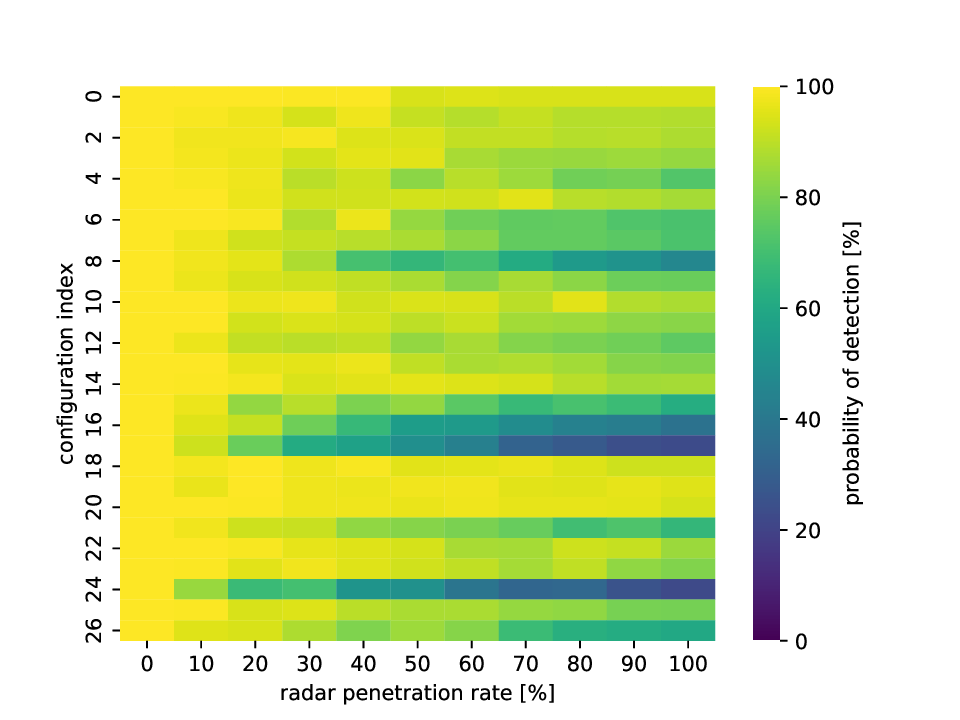}
        \caption{Predefined frequency method.}                   
        \label{fig:pd_predefined_freq}
    \end{subfigure}
    \begin{subfigure}[t]{0.24\textwidth}
        \centering 
        \includegraphics[width=\textwidth]{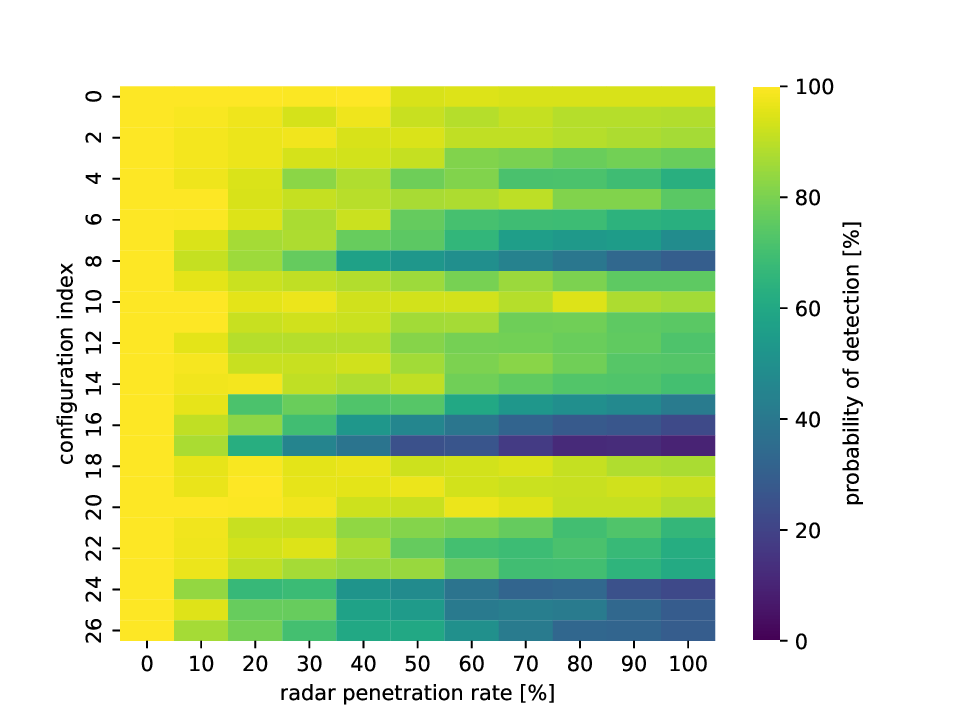}
        \caption{Predefined polarization method.}                   
        \label{fig:pd_predefined_pol}
    \end{subfigure}
    \begin{subfigure}[t]{0.24\textwidth}
    \centering 
    \includegraphics[width=\textwidth]{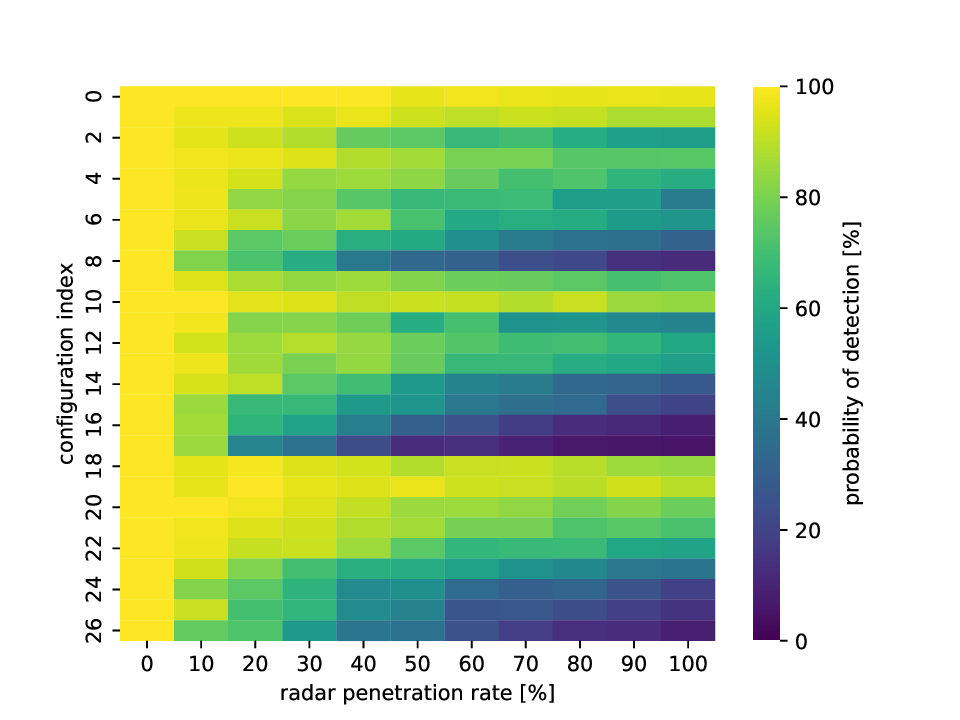}
    \caption{Time dithering method.}                   
    \label{fig:pd_time_dithering}
    \end{subfigure}
        \begin{subfigure}[t]{0.24\textwidth}
        \centering 
        \includegraphics[width=\textwidth]{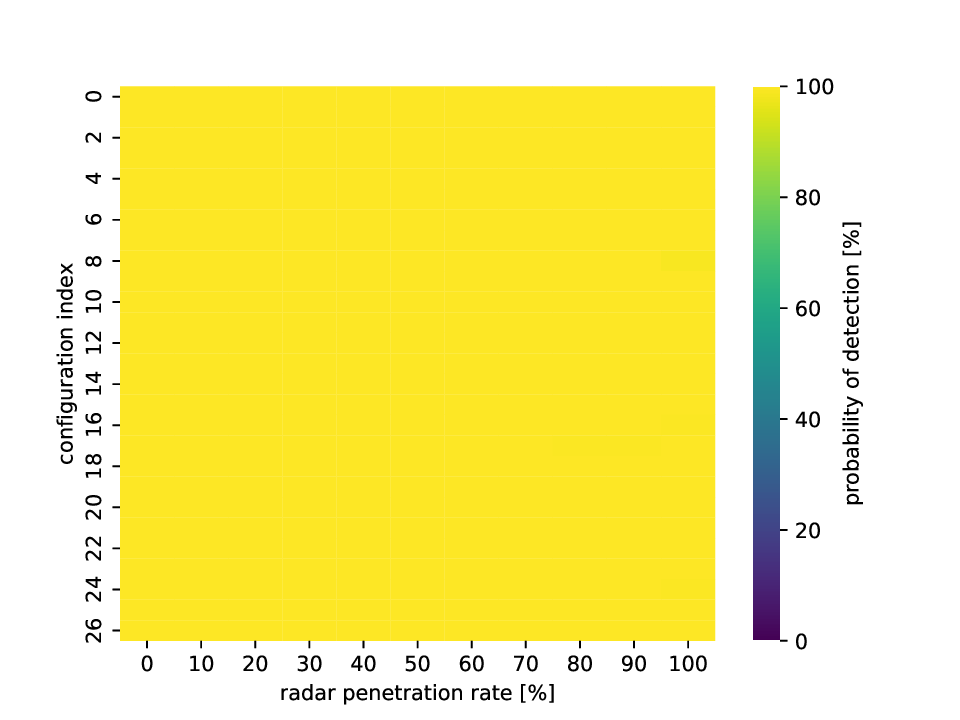}
        \caption{Time-frequency channels method.}                   
        \label{fig:pd_ofdm}
    \end{subfigure}
    \caption{Detection performance (PD) as a function of radar penetration rate under different interference mitigation techniques.}
    \label{fig:analysis_pd}
\end{figure}

Fig. \ref{fig:pd_ofdm} presents the performance of the time-frequency coding method, which consistently outperforms all other techniques. It maintains high PD levels even under high radar penetration rates, demonstrating superior robustness through structured synchronization and efficient spectrum utilization.

Among the evaluated mitigation strategies, the time-frequency coding technique provides the most effective interference mitigation, sustaining high detection performance under challenging conditions. Its structured and synchronized approach ensures effective spectrum sharing, meeting the performance requirements for advanced ADAS and AV systems. While other techniques offer partial improvements and they fail to sustain adequate radar performance in dense interference scenarios.
\begin{table}
    \centering
    \tiny
    \caption{Interference simulation configuration}
    \label{tab:configuration}
    \begin{tabular}{|c|c|c|c|}
        \hline
        Index & Scenario & Installation Topology & Host Radar Type \\
        \hline
        0 &  &  & SBZA \\
        \cline{1-1} \cline{4-4}
        1 &   & Front & SRR \\
        \cline{1-1} \cline{4-4}
        2 &   &   & LRR\\
        \cline{1-1} \cline{3-4}
        3 &  &  & SBZA  \\
        \cline{1-1} \cline{4-4}
        4 & Low & Partial & SRR \\
        \cline{1-1} \cline{4-4}
        5 &  &  & LRR\\
        \cline{1-1} \cline{3-4}
        6 &  &  & SBZA  \\
        \cline{1-1} \cline{4-4}
        7 &  & Full & SRR  \\
        \cline{1-1} \cline{4-4}
        8 &  &  & LRR \\
        \cline{1-2} \cline{3-4}
        9 &  &  & SBZA  \\
        \cline{1-1} \cline{4-4}
        10 &  & Front & SRR  \\
        \cline{1-1} \cline{4-4}
        11 &  &  & LRR \\
        \cline{1-1} \cline{3-4}
        12 &  &  & SBZA  \\
        \cline{1-1} \cline{4-4}
        13 & Medium & Partial & SRR  \\
        \cline{1-1} \cline{4-4}
        14 &  &  & LRR \\
        \cline{1-1} \cline{3-4}
        15 &  &  & SBZA  \\
        \cline{1-1} \cline{4-4}
        16 &  & Full & SRR  \\
        \cline{1-1} \cline{4-4}
        17 &  &  & LRR \\
        \cline{1-2} \cline{3-4}
        18 &  &  & SBZA  \\
        \cline{1-1} \cline{4-4}
        19 &  & Front & SRR  \\
        \cline{1-1} \cline{4-4}
        20 &  &  & LRR \\
        \cline{1-1} \cline{3-4}
        21 &  &  & SBZA  \\
        \cline{1-1} \cline{4-4}
        22 & High & Partial & SRR  \\
        \cline{1-1} \cline{4-4}
        23 &  &  & LRR \\
        \cline{1-1} \cline{3-4}
        24 &  &  & SBZA  \\
        \cline{1-1} \cline{4-4}
        25 &  & Full & SRR  \\
        \cline{1-1} \cline{4-4}
        26 &  &  & LRR \\
        \hline
    \end{tabular}
\end{table}
    
\section{Experiment}\label{sec:experiment}
Interference experiments were conducted with two primary objectives: validating the interference simulation and evaluating the effectiveness of existing interference mitigation techniques using operational radars. Two experimental environments were considered: an anechoic chamber, enabling controlled measurements, and an outdoor setup designed to emulate real-world conditions and assess simulation fidelity. 

\begin{figure}
    \centering
    \includegraphics[width=0.4\textwidth]{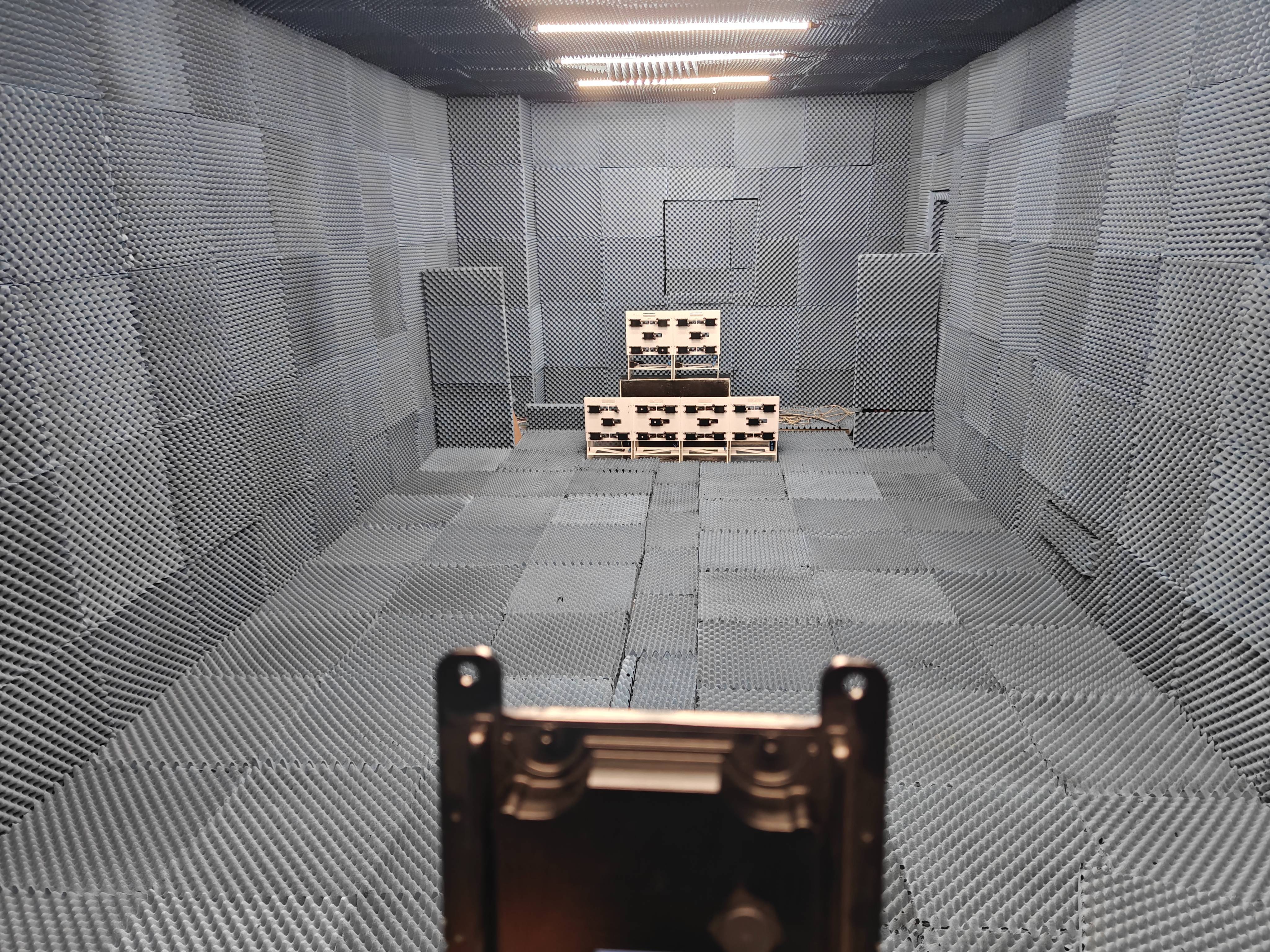}
    \caption{Anechoic chamber interference radar layout.}
    \label{fig:anechoic_layout}
\end{figure}

In the anechoic chamber, thirty interference radars were positioned $7$m from the host radar, alongside a $10$dBsm corner reflector serving as a reference target (Fig. \ref{fig:anechoic_layout}). The interference setup consisted of six arrays, each comprising five ultra-short range radars (USRR), for a total of 30 interferers. The experiments employed waveforms consistent with the simulation settings.

The tests involved three radars-under-test: A, B, and C. Radar A is an engineering sample capable of recording raw analog-to-digital converter (ADC) samples, enabling in-depth analysis of interference signals and phenomenology \cite{bilik2018automotive}. This capability supports the development of mitigation techniques applied prior to detection. Radars B and C are advanced systems with LRR and SRR capabilities, currently under evaluation, and provide detection-level outputs only. The waveform parameters of Radar A are listed in Table \ref{tab:radarA}, where $f_s$ denotes the ADC sampling rate.

\begin{table}
    \centering
    \caption{Radar A waveform}
    \label{tab:radarA}
    \begin{tabular}{|c|c|c|}
        \hline
        Parameter & Value & Units \\
        \hline
        PRI & 27.4 & $\mu s$ \\
        Slope & 26 & $MHz/\mu s$\\
        $f_s$ & 25 & $MHz$\\
        $T_c$ & 18.88 & $\mu s$\\
        $f_c$ & 76.889 & $GHz$\\
        $\#$Chirps & 512 &\\
        FPS & 15 & $Hz$ \\
        \hline
    \end{tabular}
\end{table}

\subsection{Simulation Validation}
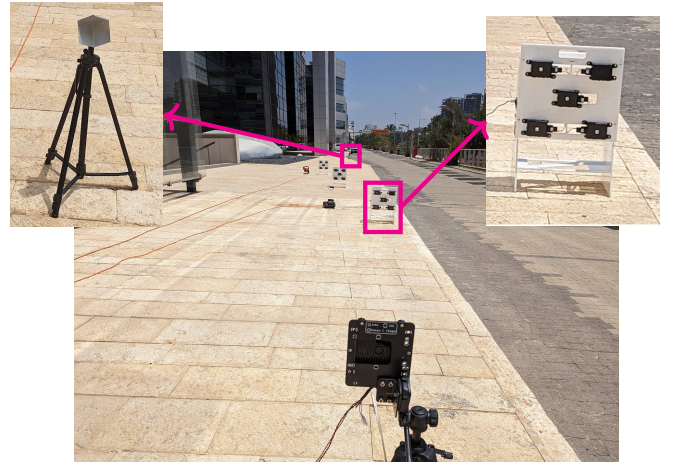
\begin{figure}
    \centering
    \begin{tikzpicture}
        \node (img1) [anchor=south west, inner sep=0pt] at (0,0) {\includegraphics[scale=.05]{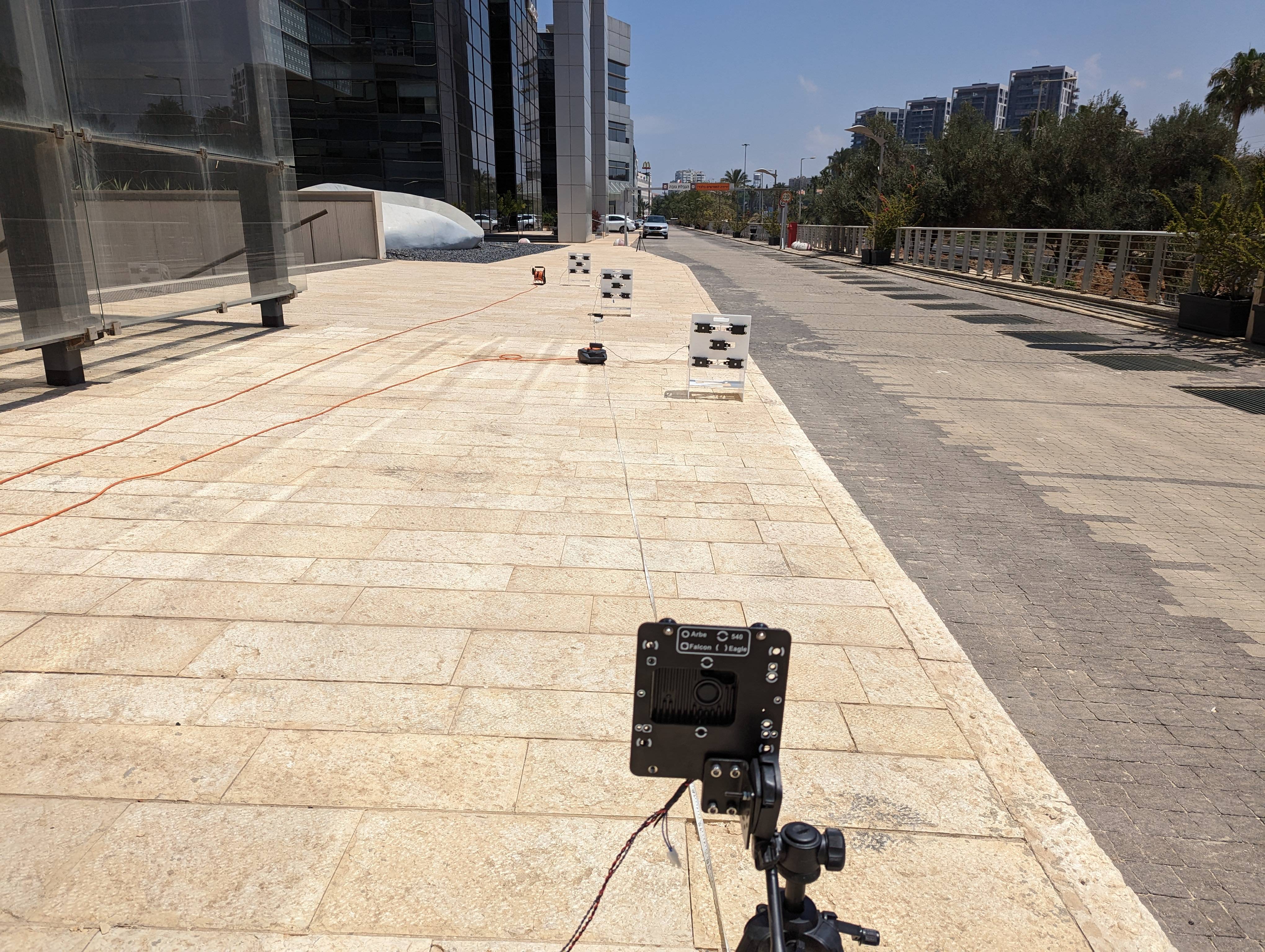}};
        \begin{scope} [x={(img1.south east)}, y={(img1.north west)}]
            \draw [magenta, ultra thick] (.4925,.720) rectangle +(.032,.05);
            \draw [magenta, ultra thick] (.535,.56) rectangle +(.065,.120);
            \node (img2) [anchor=south east, inner sep=0pt] at (1.075,.575) {\includegraphics[scale=.20]{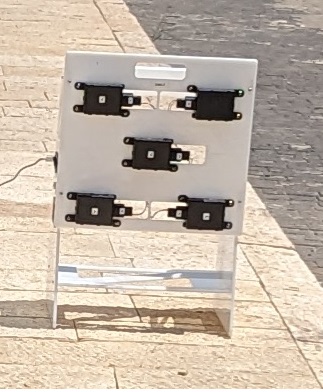}};
            \node (img3) [anchor=south east, inner sep=0pt] at (0.165,.572) {\includegraphics[scale=.3]{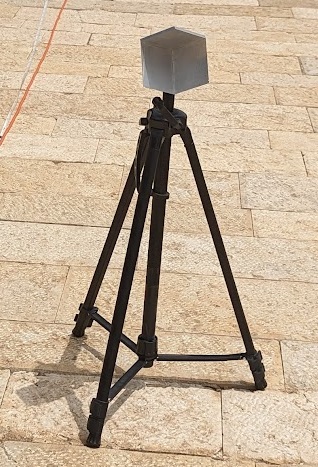}};
            \draw[magenta, ultra thick, ->] (.6, .62) -- (img2.west);
            \draw[magenta, ultra thick, ->] (.4925, .745) -- (img3.east);
        \end{scope}
        
    \end{tikzpicture}
    \captionof{figure}{Radar A field experiment layout.}
    \label{fig:cascaded_exp_layout}
\end{figure}

To validate the simulation, an equivalent field scenario was constructed using identical waveform configurations. The setup employed Radar A as the host radar-under-test, along with $15$ interference radars (three arrays) and three $10$dBsm corner reflectors as reference targets, as shown in Fig. \ref{fig:cascaded_exp_layout}. 

The interference arrays were placed at distances of $5$m, $10$m, and $15$m, from the host radar. These distances emulate long-range radar (LRR) interferers at $55$m, $110$m, and $165$m, respectively, , assuming an effective radiated power (ERP) of $35$dBm. Lower-power radars at short distances were thus used to replicate high-power distant interferers.

The locations were derived according to:
\begin{equation}
    R_{field} = R_{sim}\sqrt{\frac{P_{field}}{P_{sim}}},
\end{equation} 
where $R_{field}$ and $R_{sim}$ denote the field and simulated distances, respectively, and $P_{field}$ and $P_{sim}$ represent the corresponding radar transmit power. The waveform characteristics of the $15$ interference radars are identical to the simulation.

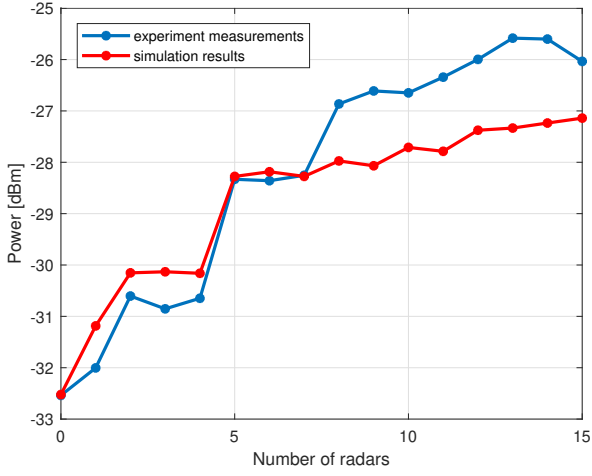
\begin{figure}
    \centering
    \begin{tikzpicture}
        \node (img1) [anchor=south west, inner sep=0pt] at (0,0) {\includegraphics[scale=.60]{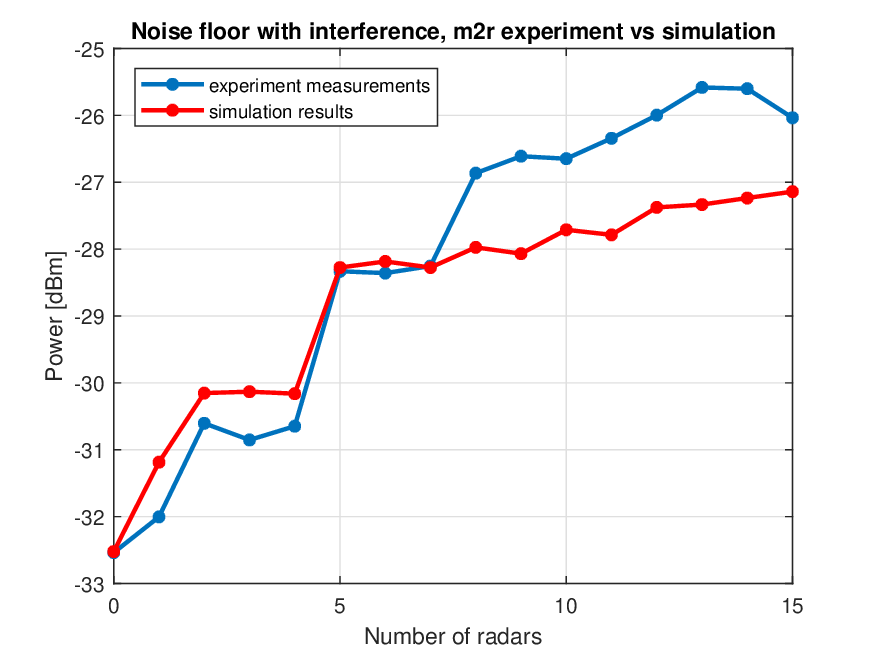}};
        \begin{scope} [x={(img1.south east)}, y={(img1.north west)}]
            \draw [white, ultra thick] (.135,0.95) rectangle +(.95,.005);
            \draw [white, ultra thick] (.135,0.945) rectangle +(.95,.00510);
            \draw [white, ultra thick] (.135,0.94) rectangle +(.95,.00510);
            \draw [white, ultra thick] (.135,0.93) rectangle +(.95,.00510);
            \draw [white, ultra thick] (.135,0.935) rectangle +(.95,.00510);
            \draw [white, ultra thick] (.135,0.96) rectangle +(.95,.00510);
        \end{scope}
    \end{tikzpicture}
    \captionof{figure}{Average noise floor versus number of active interferers. The blue curve shows experimental results, and the red curve shows simulation results.}
    \label{fig:nf_sim_vs_exp_poster}
\end{figure}

The interference radars were activated incrementally. The resulting average noise floor is shown by the blue line in Fig. \ref{fig:nf_sim_vs_exp_poster}. A total increasenof $6.5dB$ was observed for $15$ active interferers. As a result, the radar maximum detection range experienced a reduction of $31\%$, according to the radar equation \cite{skolnik}:
\begin{equation}
    R_{max} = \sqrt[4]{\frac{P_tG_tG_r\lambda^2}{(4\pi)^3 SNR_{min}}} \;.
\end{equation}

The experimental results closely match the simulation, with deviations of up to $1.5$dB. These discrepancies are attributed to hardware non-idealities, environmental variability, and modeling assumptions, confirming the validity of the simulation framework.

\subsection{Noise Level Rise}
As mentioned before, interference primarily manifests as an increase in the noise floor. This effect was evaluated across different radars and interference levels in both anechoic and field environments..

\begin{figure}
    \centering
    \begin{subfigure}[t]{0.24\textwidth}
        \centering
        \includegraphics[width=\textwidth,trim={1mm 1mm 1mm 1mm},clip]{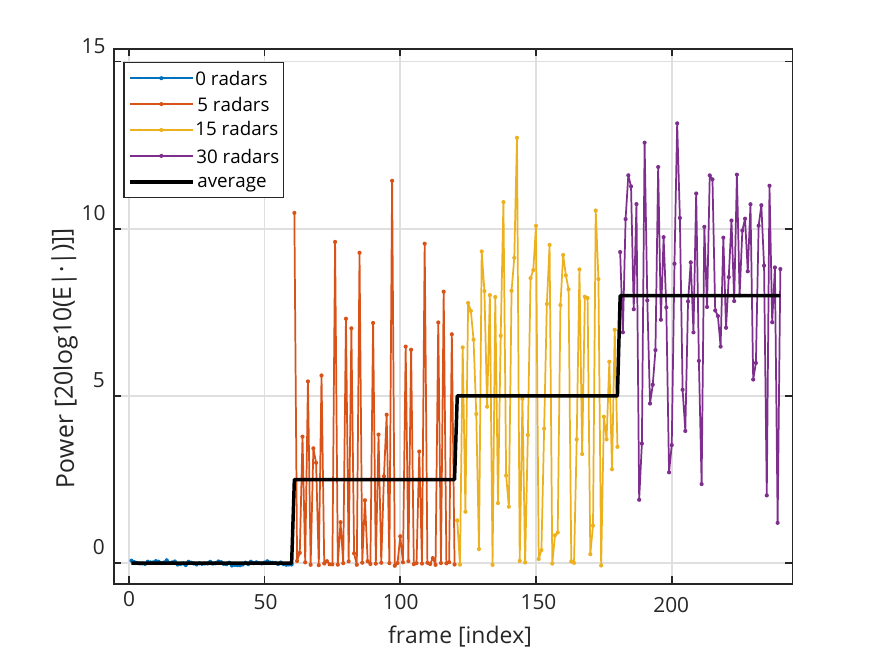}
    \caption{Radar A noise floor.}
    \label{fig:m2r_all_normalized}
    \end{subfigure}
    \centering
    \begin{subfigure}[t]{0.24\textwidth}
        \centering
        \includegraphics[width=\textwidth,trim={1mm 1mm 1mm 1mm},clip]{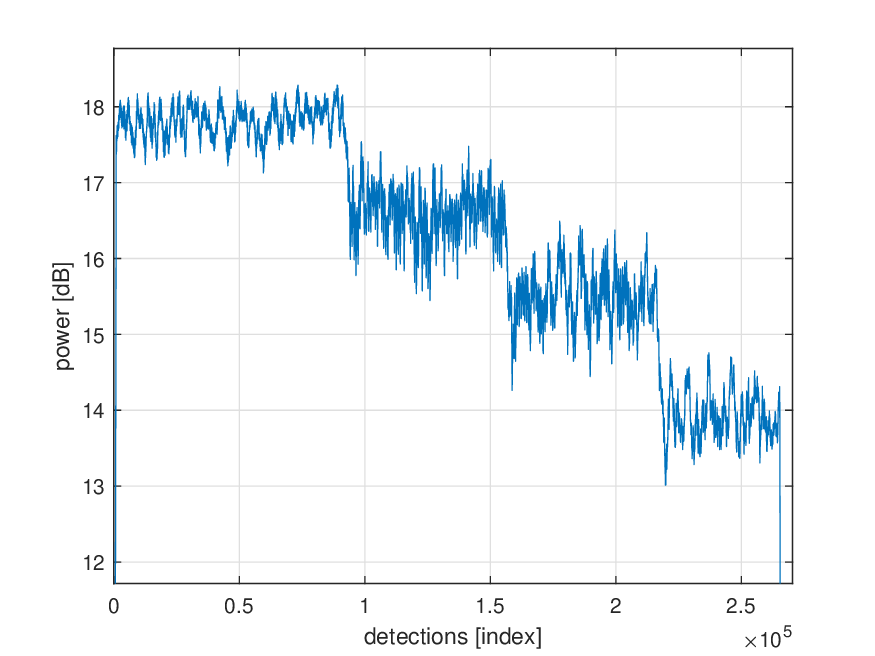}
        \caption{Radar B SNR.}
        \label{fig:snr_moving_average_hella_all}
    \end{subfigure}
    \caption{Noise floor and SNR degradation with increasing interference in an anechoic chamber.}
    \label{fig:NoiseLvlRise}
\end{figure}

First, radars A and B were tested in an anechoic chamber with $5$, $15$, and $30$ active interferers.
For Radar A, Fig. \ref{fig:m2r_all_normalized} shows a progressive noise floor increase of $2.5$dB, $5$dB, and $8$dB, respectively. This increase reduces the signal-to-noise ratio (SNR) and consequently degrades detection performance \cite{richards2005fundamentals}..

Radar B, which outputs a detection-level SNR values, was also tested under similar conditions. Fig. \ref{fig:snr_moving_average_hella_all} shows that the reference target SNR decreases by approximately $1$dB, $2.5$dB, and $4$dB for $5$, $15$, and $30$ interferers, respectively, consistent with the observations from Radar A.

Despite identical interference conditions, the two radars exhibit different sensitivities. This can be attributed to variations in their system designs, including differences in antennas, receivers, and signal processing flows.

\begin{figure}
    \centering
    \includegraphics[width=0.4\textwidth]{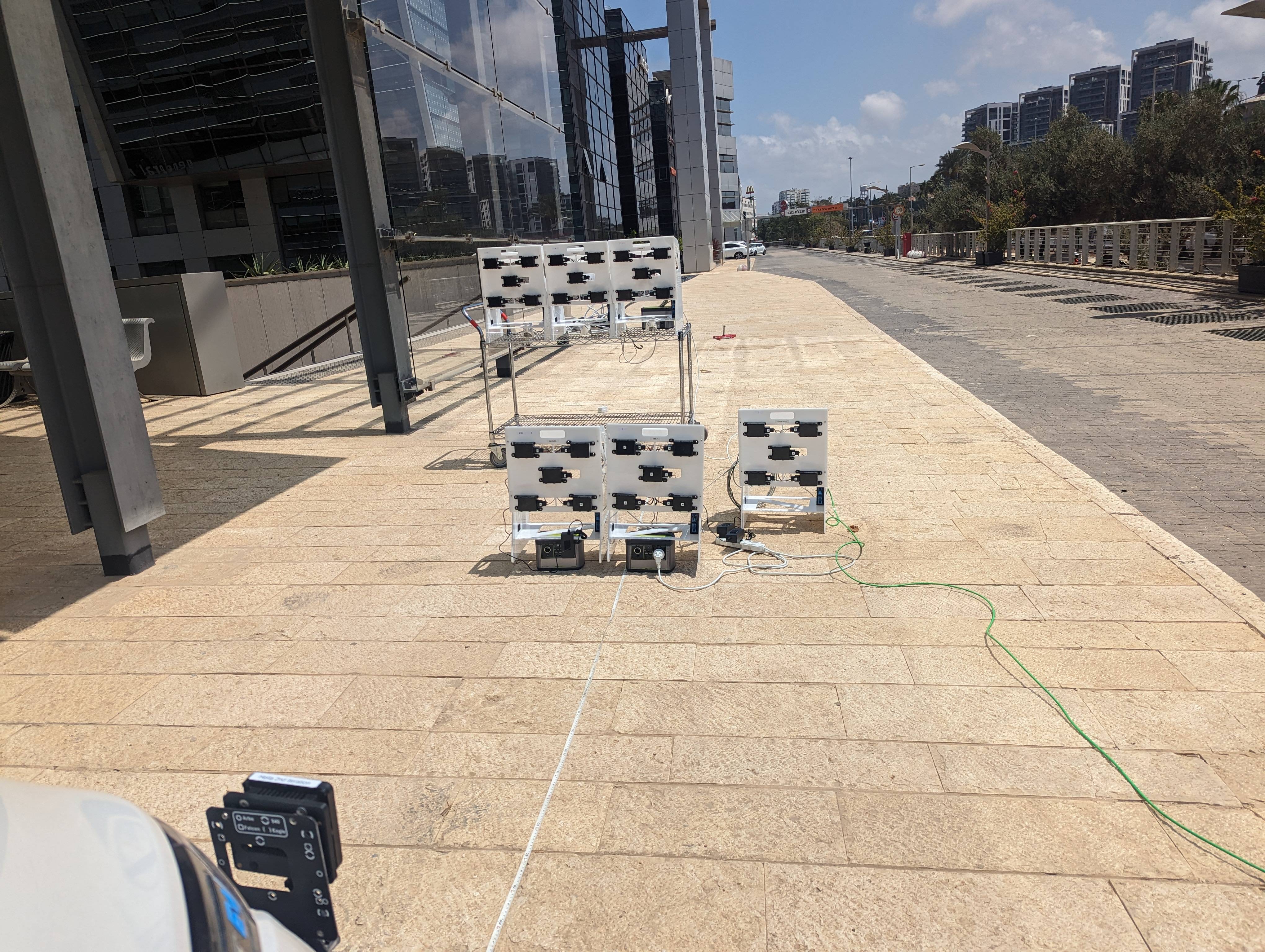}
    \caption{Radar B field experiment layout.}\label{fig:hella_field_experiment}
\end{figure}

A field experiment with Radar B (Fig. \ref{fig:hella_field_experiment}) further demonstrates these effects. This experiment features $30$ interference radars and corner reflectors as reference targets. The waveforms of the first $20$ activated interference radars are identical to the simulation. The remaining $10$ interference radars employ waveforms similar to that of Radar B. While Radar B utilizes a stepped-FM waveform, these interference radars do not implement stepped-FM \cite{levanon2004radar}.

\begin{figure}
    \centering
    \includegraphics[width=0.5\textwidth]{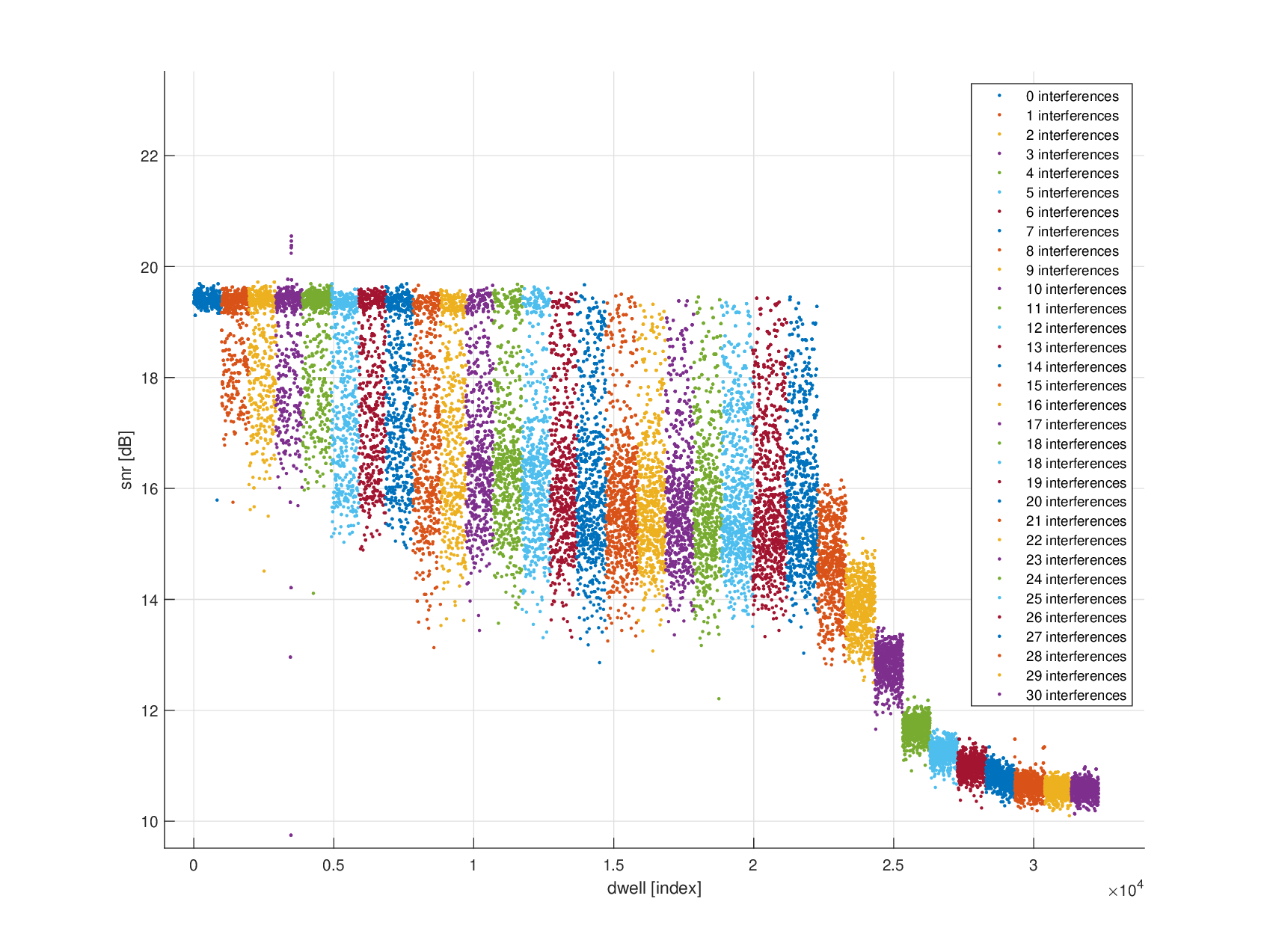}
    \caption{SNR value from the corner reflector at 60m with radar B, with increasing number of interference radars.}\label{fig:cr_mixed}
\end{figure}

Fig. \ref{fig:cr_mixed} presents the SNR measurements of a corner reflector located $60$m from the radar-under-test. In the absence of interference, the SNR of the corner reflector is $19.5dB$. As interference radars are incrementally activated, the SNR decreases. Where the significant SNR drop occurs when radars $21-30$, which have a similar waveform, were activated. This similarity results in more severe interference and greater performance degradation.

\begin{figure}
    \centering
    \includegraphics[width=0.5\textwidth]{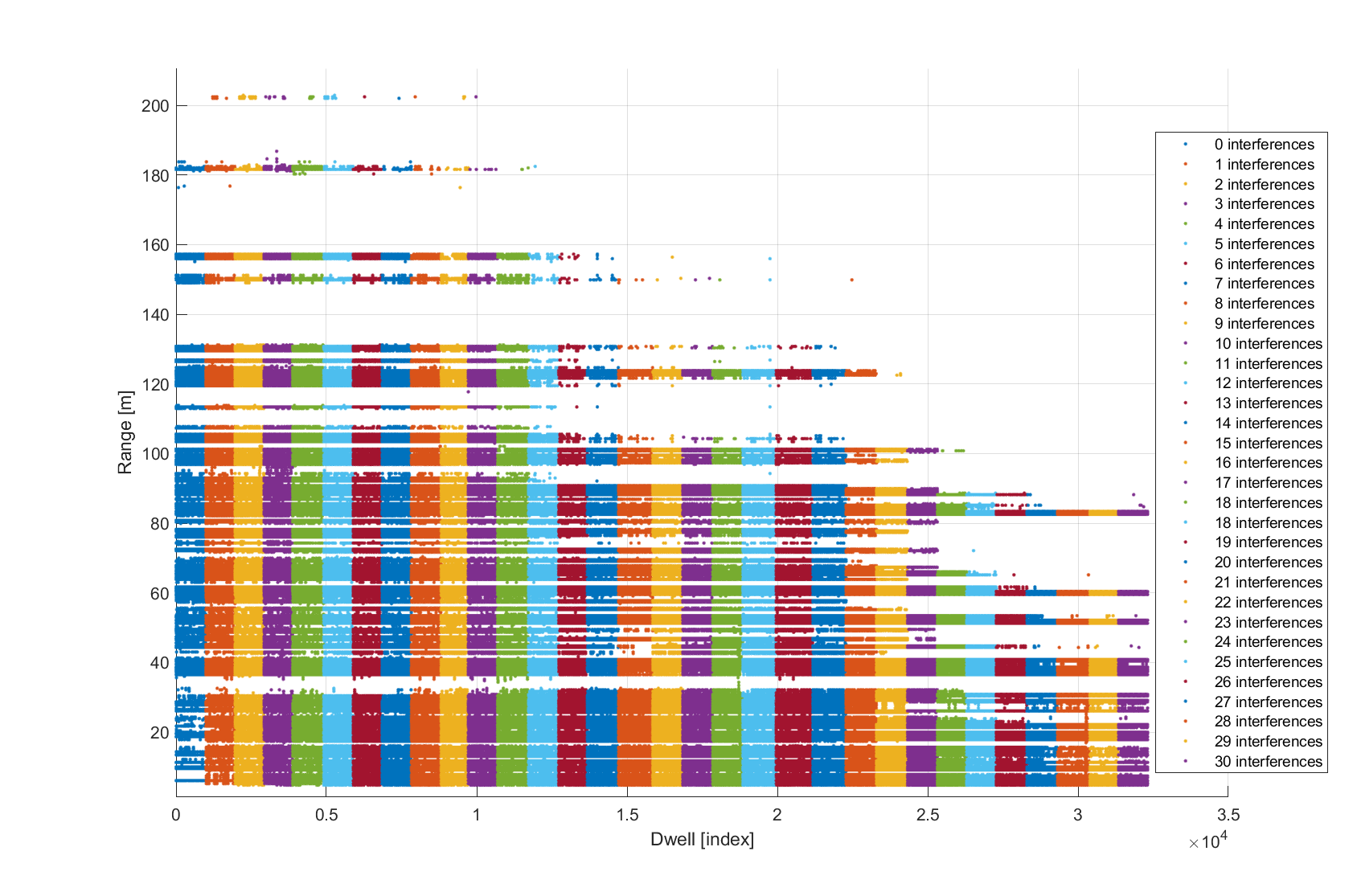}
    \caption{Maximal detection range degradation with increasing interference for victim radar B.}\label{fig:detection_ranges_mixed}
\end{figure}

The rise in noise floor level and therefore the decrease in SNR reduce the maximal detection range (Fig. \ref{fig:detection_ranges_mixed}). As more interference radars become active, the ability to detect distant and faint targets diminishes. A target at $180$m begins to show degraded detection performance once $7$ interference radars are active, and it becomes undetectable after $12$ interference radars are turned on. In this scenario, the maximum detection range for targets in this scenario is reduced from $200$m to $80$m, illustrating the significant impact of interference on Radar B’s performance.

\begin{figure}
    \centering
    \begin{subfigure}[t]{0.22\textwidth}
        \centering
        \includegraphics[width=\textwidth]{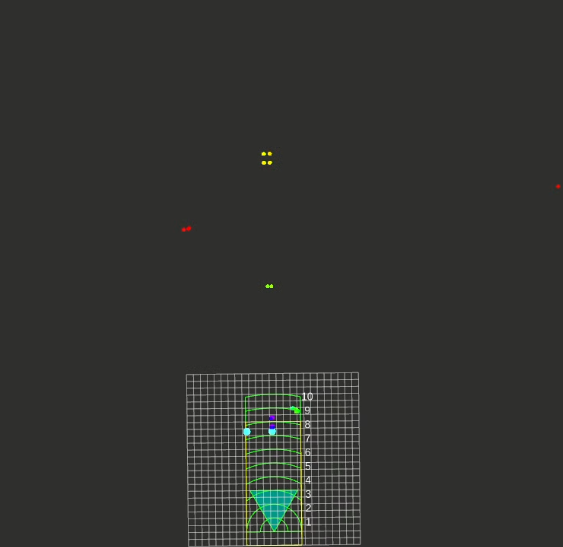}
        \caption{Fixed threshold detector without interference.}
        \label{fig:radar_b_no_interference}
    \end{subfigure}
    \begin{subfigure}[t]{0.22\textwidth}
        \centering 
        \includegraphics[width=\textwidth]{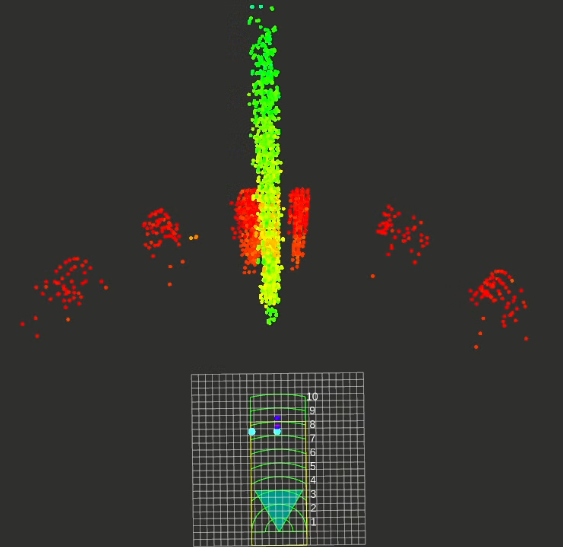}
        \caption{Fixed threshold detector with interference, many false detection are detected.}                   
        \label{fig:radar_b_with_interference}
    \end{subfigure}
    \begin{subfigure}[t]{0.22\textwidth}
        \centering 
        \includegraphics[width=\textwidth]{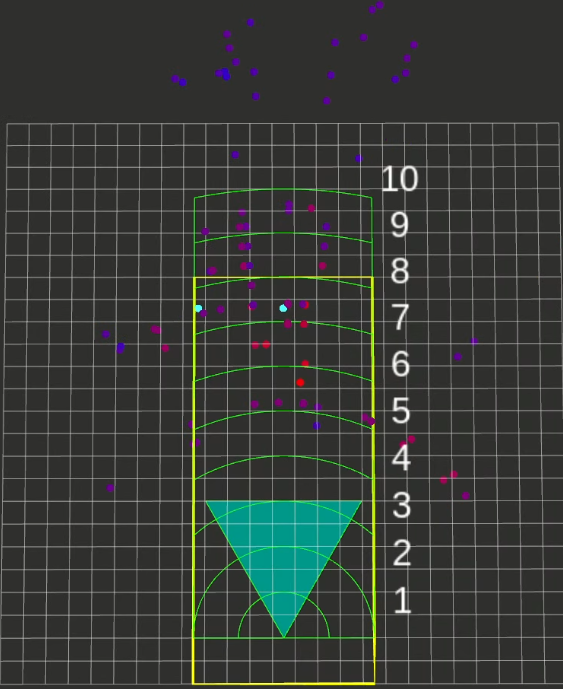}
        \caption{CA-CFAR detector without interference.}                   
        \label{fig:hella_occulii_no_interference}
    \end{subfigure}
    \begin{subfigure}[t]{0.22\textwidth}
        \centering 
        \includegraphics[width=\textwidth]{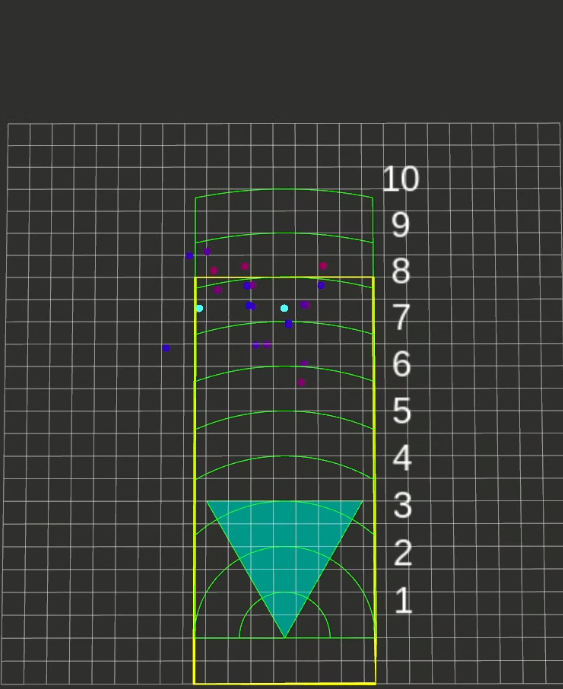}
        \caption{CFAR detector with interference, weak detections are not detected.}                   
        \label{fig:hella_occulii_with_interference}
    \end{subfigure}
    \caption{Interference effects on fixed threshold and CA-CFAR detectors in anechoic chamber experiments.}
    \label{fig:detector_with_interference}
\end{figure}

The increase in the noise level might result in a high number of false detections, while target detection is not affected, or a reduction in PD, due to reduction in SNR values, as was shown previously. The detector that is used in the processing chain influence on the type of phenomenon it will face. Radar B utilizes a CA-CFAR detector, while radar C, has a fixed threshold detector. They were evaluated in a anechoic chamber experiment with and without interference. Fig. \ref{fig:detector_with_interference} shows a top view of the evaluated scenario, where each point indicates a detection and its color its SNR value (red color is higher SNR than blue color). Subfigures (a) and (b) show Radar C detections with and without interference, and (c) and (d) show Radar B detections with and without interference, respectively.

Subfigure (a) shows only a few detections originating from objects in the chamber are visible. When interference is introduced, multiple false detections surpass the fixed threshold. These false detections appear in two geometries: one along the range axis in the radar boresight (mostly green) and another following a curved line (mostly red). The false alarms significantly impair Radar C's ability to serve as a reliable perception sensor.

In contrast, Radar B, demonstrates superior performance in the presence of interference. In subfigure (c) several detections can be seen, in the absence of interference, while subfigure (d) reveals fewer detections due the activation of interference with reduced SNR due to the increased noise, as indicated by their color. Notably, subfigure (d) is devoid of the false alarms observed in Radar C's experiment. As the noise floor rises, the adaptive threshold of the CA-CFAR detector also increases, effectively preventing the detection of the elevated noise floor, which was responsible for the false alarms observed in Radar C. 

\subsection{Influence on Processing Chain}
\begin{figure}
    \centering
    \begin{subfigure}[t]{0.22\textwidth}
        \centering
        \includegraphics[width=\textwidth]{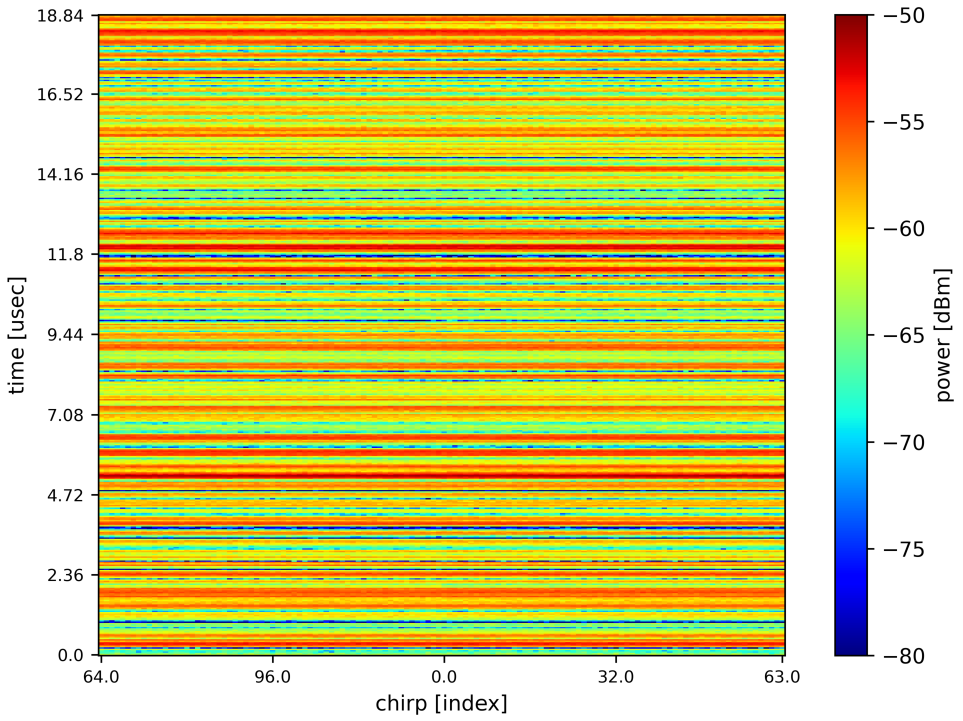}
        \caption{Without after Sampling.}
        \label{fig:time_map_no_mitigation_no_interference}
    \end{subfigure}
    \begin{subfigure}[t]{0.22\textwidth}
        \centering
        \includegraphics[width=\textwidth]{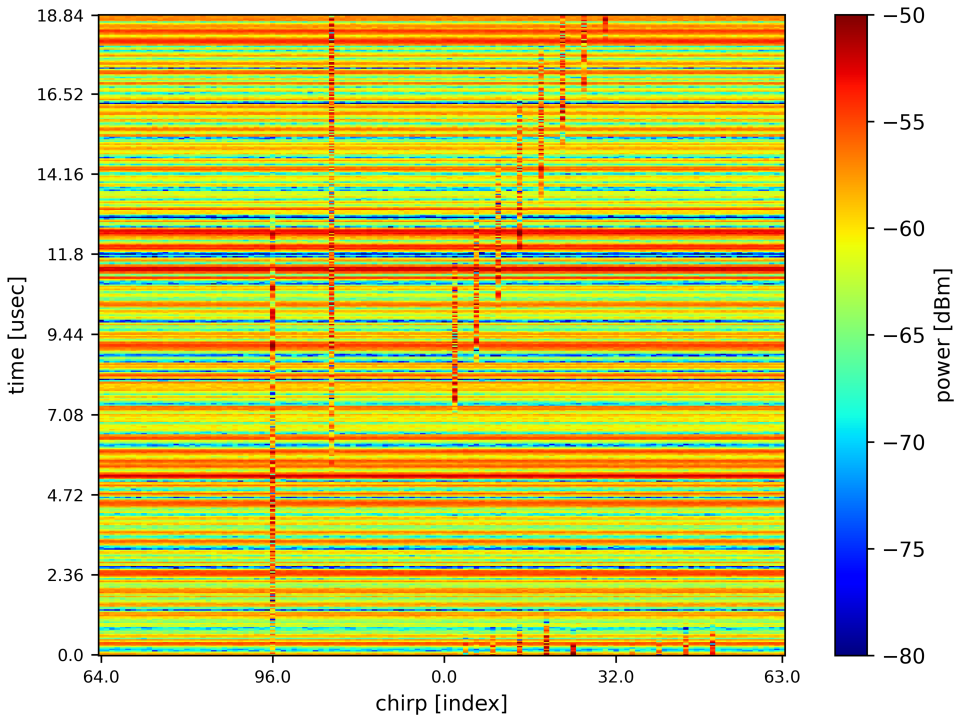}
        \caption{With after Sampling.}
        \label{fig:time_map_no_mitigation_with_interference}
    \end{subfigure}
    \begin{subfigure}[t]{0.22\textwidth}
        \centering
        \includegraphics[width=\textwidth]{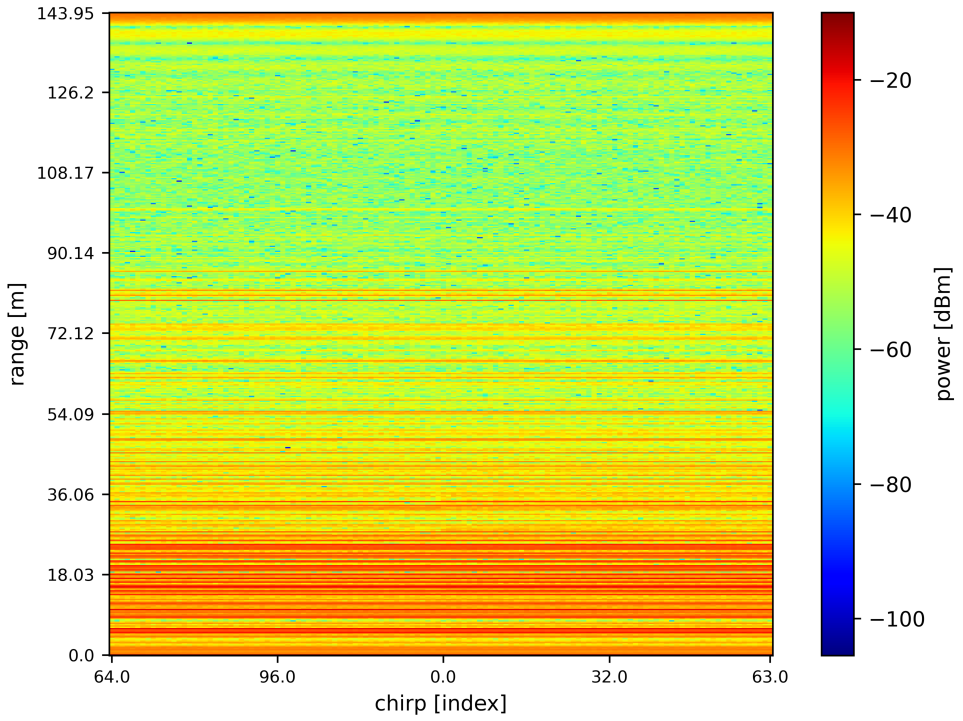}
        \caption{Without after range FFT.}
        \label{fig:range_map_no_mitigation_no_interference}
    \end{subfigure}
    \begin{subfigure}[t]{0.22\textwidth}
        \centering
        \includegraphics[width=\textwidth]{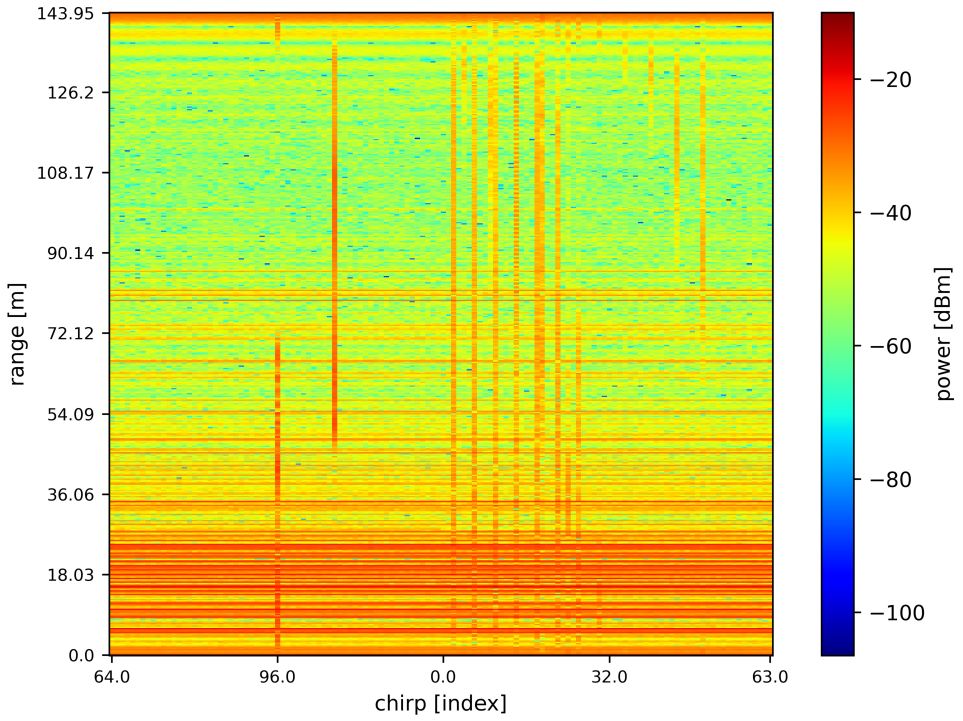}
        \caption{With after range FFT.}
        \label{fig:range_map_no_mitigation_with_interference}
    \end{subfigure}
    \begin{subfigure}[t]{0.22\textwidth}
        \centering
        \includegraphics[width=\textwidth]{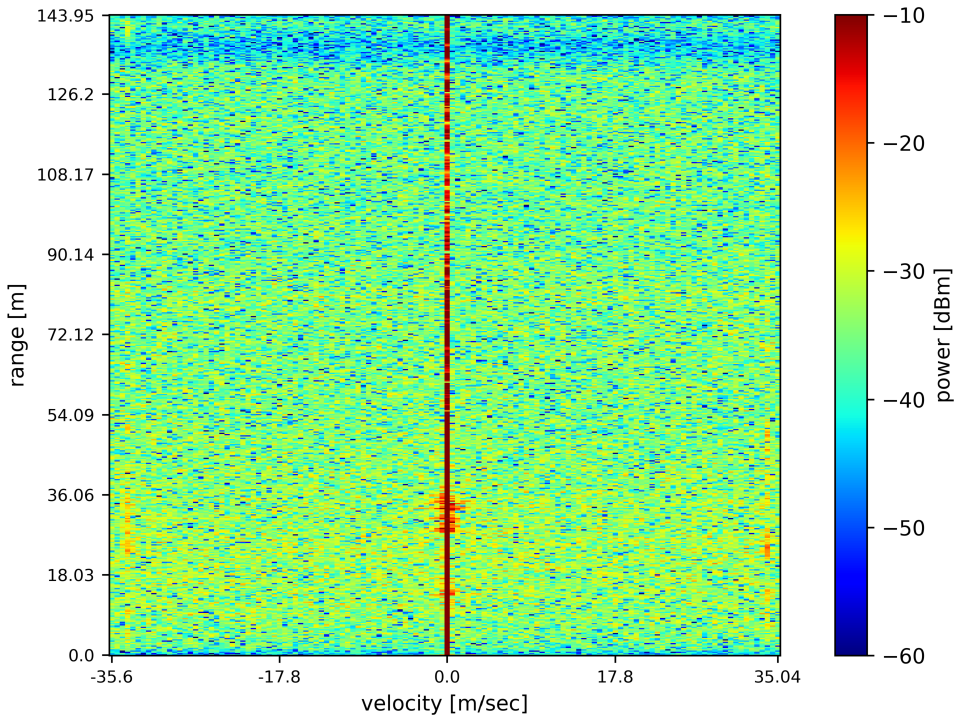}
        \caption{Without after Doppler FFT.}
        \label{fig:doppler_map_no_mitigation_no_interference}
    \end{subfigure}
    \begin{subfigure}[t]{0.22\textwidth}
        \centering
        \includegraphics[width=\textwidth]{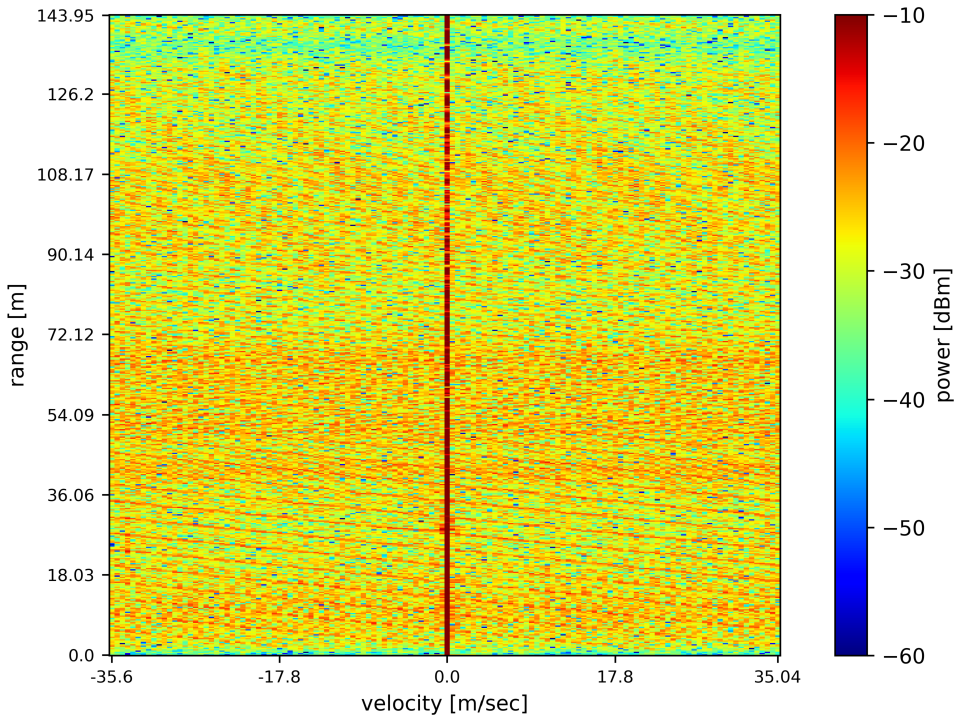}
        \caption{With after Doppler FFT.}
        \label{fig:doppler_map_no_mitigation_with_interference}
    \end{subfigure}
    \caption{Signal maps at different processing stages with and without interference.}
    \label{fig:Signalmaps}
\end{figure}

Access to ADC samples enables detailed analysis of interference throughout the radar-under-test processing chain. Fig. \ref{fig:Signalmaps} presents the output maps at various stages of the radar system processing flow (\ref{fig:radar_system}.) after ADC, after range FFT and after Doppler FFT. The maps show the radar at two states, with and without interference.
Interference appears as vertical structures in the time–chirp (Fig. \ref{fig:time_map_no_mitigation_with_interference}) and range–chirp (Fig. \ref{fig:range_map_no_mitigation_with_interference}) domains, since each vertical line represents a chirp, and the interference affects a portion of the chirp. In the time-chirp map, the affected bins correspond to samples where interference is received, while in the range-chirp map, the affected samples represent frequencies where interference manifests. After Doppler processing (Fig. \ref{fig:doppler_map_no_mitigation_with_interference}), these effect appear as horizontal noise bands, corresponding to the previously observed vertical lines. These regions are smeared in the Doppler domain due to the FFT process. In this experiment, five interference radars (a single array) were activated.

\section{Conclusions} \label{sec:con}
This paper presented a comprehensive analysis of automotive radar interference through simulation and experimental validation. The results demonstrate substantial performance degradation as interference levels increase, driven by higher radar penetration rates and growing radar density per vehicle. Notably, even at relatively low penetration rates, radar performance may fall below operational requirements in the absence of effective mitigation strategies.

Conventional interference mitigation techniques provide limited robustness and are effective primarily in low-interference environments. However, their performance degrades rapidly as radar density increases. The results indicate that advanced mitigation approaches are required to sustain reliable radar operation under realistic, high-density conditions. In particular, the findings highlight the importance of coordinated and collaborative strategies among radar systems to enable efficient spectrum utilization and mitigate mutual interference.

The simulation framework was validated through controlled and field experiments involving up to $30$ interfering radars. The strong agreement between simulation and experimental results confirms the accuracy of the proposed modeling approach. The experiments further demonstrate the high susceptibility of current radar systems to interference, with significant degradation in detection performance under dense interference conditions.

Overall, this work underscores the critical need for robust interference mitigation techniques and cooperative spectrum management mechanisms to support the reliable deployment of automotive radar systems in future dense traffic environments. Future work will focus on the development and evaluation of collaborative interference mitigation schemes.

\bibliographystyle{IEEEtran}
\bibliography{References.bib}	

\end{document}

%% file: Definitions.tex
\documentclass[journal,twoside]{IEEEtran}

\usepackage{cite}
\usepackage{mathtools}
\usepackage{bm}
\usepackage{amsmath,amssymb,amsfonts}
\usepackage{graphicx}
\usepackage{textcomp}
\usepackage{bm}
\usepackage{color}
\usepackage{epstopdf}
\usepackage{subcaption}
\usepackage{enumitem}
\usepackage{tikz}
\usetikzlibrary{shapes.geometric, arrows}
\usetikzlibrary{positioning}
\usepackage{algorithm}
\usepackage{algpseudocode}
\usepackage{color, colortbl}
\definecolor{Gray}{gray}{0.1}
\usepackage{tikz}
\usepackage{titlesec}
\usepackage{xcolor}
\usepackage{soul}

\usepackage{stfloats}
\usepackage{hyperref}
\hypersetup{colorlinks=true,citecolor=blue,linkcolor=blue,urlcolor=blue}
\usepackage{cleveref}
\usepackage{caption}

\usepackage{tikz}
\usetikzlibrary{shapes.geometric, arrows}
\usetikzlibrary{circuits.ee.IEC}
\usetikzlibrary{positioning}
\tikzstyle{arrow} = [thick,->,>=stealth]
\tikzstyle{arrowDashed} = [thick,->,>=stealth,dashed]
\tikzstyle{GrayBox} = [rectangle, minimum height=1.2cm, text centered, draw=black, fill=gray!20!,rounded corners]
\tikzstyle{YellowBox} = [rectangle, minimum height=1.2cm, text centered, draw=black, fill=yellow!50!,rounded corners]
\tikzstyle{GreenBox} = [rectangle, minimum height=1.2cm, text centered, draw=black, fill=lime!50!,rounded corners]
\tikzstyle{BlueBox} = [rectangle, minimum height=1.2cm, text centered, draw=black, fill=cyan!50!,rounded corners]
\tikzstyle{RedBox} = [rectangle, minimum height=1.2cm, text centered, draw=black, fill=red!10!,rounded corners]
\tikzstyle{EmptyBox} = [rectangle, minimum width=1cm, minimum height=0.2cm, text centered]

\usepackage{tikz}
\usetikzlibrary{shapes.geometric, arrows}
\usetikzlibrary{circuits.ee.IEC}
\usetikzlibrary{positioning}
\usetikzlibrary{shapes, shadows}
\usetikzlibrary{positioning,shapes,arrows.meta}

\tikzstyle{arrow} = [thick,->,>=stealth]
\tikzstyle{FullBox} = [rounded corners, minimum height=1cm, text centered, draw=black,fill=blue!10!]
\tikzstyle{EmptyBox} = [rectangle, minimum width=1cm, minimum height=0.2cm, text centered]
\tikzstyle{EmptyBox1} = [rectangle, inner sep=0,outer sep=0,fill=black]
\tikzstyle{Diamond} = [diamond,aspect=1.1, minimum width=0.2cm, text centered, draw=black,fill=blue!10!]
\tikzstyle{BigBox} = [rectangle, minimum height=7cm, minimum width=5.5cm, draw=black]
\tikzstyle{Data} = [trapezium , minimum height=1cm, text centered, draw=black,trapezium left angle=60, trapezium right angle=120,fill=pink!10!]

%% file: References.bib
@article{zhang2020vanet,
  title={VANET-assisted interference mitigation for millimeter-wave automotive radar sensors},
  author={Zhang, Mengyuan and He, Shibo and Yang, Chaoqun and Chen, Jiming and Zhang, Junshan},
  journal={IEEE Network},
  volume={34},
  number={2},
  pages={238--245},
  year={2020}
}

@article{kui2021interference,
  title={Interference Analysis for mmWave Automotive Radar Considering Blockage Effect},
  author={Kui, Liping and Huang, Sai and Feng, Zhiyong},
  journal={Sensors},
  volume={21},
  number={12},
  pages={3962},
  year={2021},
  publisher={MDPI}
}

@article{brooker2007mutual,
  title={Mutual interference of millimeter-wave radar systems},
  author={Brooker, Graham M},
  journal={IEEE Transactions on Electromagnetic Compatibility},
  volume={49},
  number={1},
  pages={170--181},
  year={2007}
}

@article{schipper2015simulative,
  title={Simulative prediction of the interference potential between radars in common road scenarios},
  author={Schipper, Tom and Prophet, Silvia and Harter, Marlene and Zwirello, Lukasz and Zwick, Thomas},
  journal={IEEE Transactions on Electromagnetic Compatibility},
  volume={57},
  number={3},
  pages={322--328},
  year={2015},
  publisher={IEEE}
}

@article{al2017stochastic,
  title={Stochastic geometry methods for modeling automotive radar interference},
  author={Al-Hourani, Akram and Evans, Robin J and Kandeepan, Sithamparanathan and Moran, Bill and Eltom, Hamid},
  journal={IEEE Transactions on Intelligent Transportation Systems},
  volume={19},
  number={2},
  pages={333--344},
  year={2017}
}

@inproceedings{terbas2019radar,
  title={Radar to radar interference in common traffic scenarios},
  author={Terbas, Dilge and Laghezza, Francesco and Jansen, Feike and Filippi, Alessio and Overdevest, Jeroen},
  booktitle={2019 16th European Radar Conference (EuRAD)},
  pages={177--180},
  year={2019}
}

@article{munari2018stochastic,
  title={Stochastic geometry interference analysis of radar network performance},
  author={Munari, Andrea and Simi{\'c}, Ljiljana and Petrova, Marina},
  journal={IEEE Communications Letters},
  volume={22},
  number={11},
  pages={2362--2365},
  year={2018}
}

@article{aydogdu2020radar,
  title={Radar interference mitigation for automated driving: Exploring proactive strategies},
  author={Aydogdu, Canan and Keskin, Musa Furkan and Carvajal, Gisela K and Eriksson, Olof and Hellsten, Hans and Herbertsson, Hans and Nilsson, Emil and Rydstrom, Mats and Vanas, Karl and Wymeersch, Henk},
  journal={IEEE Signal Processing Magazine},
  volume={37},
  number={4},
  pages={72--84},
  year={2020}
}

@inproceedings{ossowska2020imiko,
  title={{IMIKO}-Radar Project: Laboratory Interference Measurements of Automotive Radar Sensors},
  author={Ossowska, Alicja and Sit, Leen and Manchala, Sarath and Vogler, Thomas and Krupinski, Kevin and Luebbert, Urs},
  booktitle={2020 21st International Radar Symposium (IRS)},
  pages={334--338},
  year={2020}
}

@article{boban2019multi,
  title={Multi-band vehicle-to-vehicle channel characterization in the presence of vehicle blockage},
  author={Boban, Mate and Dupleich, Diego and Iqbal, Naveed and Luo, Jian and Schneider, Christian and M{\"u}ller, Robert and Yu, Ziming and Steer, David and J{\"a}ms{\"a}, Tommi and Li, Jian and others},
  journal={IEEE access},
  volume={7},
  pages={9724--9735},
  year={2019}
}

@article{kurz2021road,
  title={Road Surface Characteristics for the Automotive 77 {GHz} Band},
  author={Kurz, Vera and Stuelzebach, Hannes and Pfeiffer, Florian and van Driesten, Carlo and Biebl, Erwin},
  journal={Advances in Radio Science},
  volume={19},
  number={F.},
  pages={165--172},
  year={2021},
  publisher={Copernicus GmbH}
}

@inproceedings{kamann2018automotive,
  title={Automotive radar multipath propagation in uncertain environments},
  author={Kamann, Alexander and Held, Patrick and Perras, Florian and Zaumseil, Patrick and Brandmeier, Thomas and Schwarz, Ulrich T},
  booktitle={2018 21st International Conference on Intelligent Transportation Systems (ITSC)},
  pages={859--864},
  year={2018},
  organization={IEEE}
}

@inproceedings{kunert2012eu,
  title={The {EU} project {MOSARIM}: A general overview of project objectives and conducted work},
  author={Kunert, Martin},
  booktitle={2012 9th European Radar Conference},
  pages={1--5},
  year={2012},
  organization={IEEE}
}

@article{kunert2012d5,
  title={D5. 3--Recommendations on sensor design, mounting and operational parameters to minimize radar interference},
  author={Kunert, M},
  journal={MOSARIM project report},
  year={2012}
}

@article{chipengo2018full,
  title={Full physics simulation study of guardrail radar-returns for 77 {GHz} automotive radar systems},
  author={Chipengo, Ushemadzoro},
  journal={IEEE Access},
  volume={6},
  pages={70053--70060},
  year={2018}
}

@inproceedings{sato1995measurements,
  title={Measurements of reflection characteristics and refractive indices of interior construction materials in millimeter-wave bands},
  author={Sato, Katsuyoshi and Kozima, Hideki and Masuzawa, Hiroshi and Manabe, Takeshi and Ihara, Toshio and Kasashima, Yoshinori and Yamaki, Katsunori},
  booktitle={1995 IEEE 45th Vehicular Technology Conference. Countdown to the Wireless Twenty-First Century},
  volume={1},
  pages={449--453},
  year={1995}
}

@article{3gpp2019,
  title={Study on Evaluation Methodology of New Vehicle-to-Everything {(V2X)} Use Cases for {LTE} and {NR}, (Release 15)},
  author = {{3rd Generation Partnership Project, Technical Specification Group Radio Access Network}},
  journal = {{3GPP} Technical Report {(TR)} 37.885 V15.3.0},
  year={June, 2019},
}

@article{bilik2019rise,
  title={The rise of radar for autonomous vehicles: Signal processing solutions and future research directions},
  author={Bilik, Igal and Longman, Oren and Villeval, Shahar and Tabrikian, Joseph},
  journal={IEEE signal processing Magazine},
  volume={36},
  number={5},
  pages={20--31},
  year={2019},
}

@article{solodky2020cdma,
  title={{CDMA-MIMO} radar with the tansec waveform},
  author={Solodky, Gaston and Longman, Oren and Villeval, Shahar and Bilik, Igal},
  journal={IEEE Transactions on Aerospace and Electronic Systems},
  volume={57},
  number={1},
  pages={76--89},
  year={2020}
}

@inproceedings{bilik2018automotive,
  title={Automotive multi-mode cascaded radar data processing embedded system},
  author={Bilik, Igal and Villeval, Shahar and Brodeski, Daniel and Ringel, Haim and Longman, Oren and Goswami, Piyali and Kumar, Chethan YB and Rao, Sandeep and Swami, Pramod and Jain, Anshu and others},
  booktitle={2018 IEEE Radar Conference (RadarConf18)},
  pages={0372--0376},
  year={2018}
}

@book{levanon2004radar,
  title={Radar signals},
  author={Levanon, Nadav and Mozeson, Eli},
  year={2004},
  publisher={John Wiley \& Sons}
}

@inproceedings{toth2021analysis,
  title={Analysis of automotive radar interference mitigation for real-world environments},
  author={Toth, Mate and Rock, Johanna and Meissner, Paul and Melzer, Alexander and Witrisal, Klaus},
  booktitle={2020 17th European Radar Conference (EuRAD)},
  pages={176--179},
  year={2021}
}

@inproceedings{alhumaidi2021interference,
  title={Interference avoidance and mitigation in automotive radar},
  author={Alhumaidi, M and Wintermantel, M},
  booktitle={2020 17th European Radar Conference (EuRAD)},
  pages={172--175},
  year={2021},
  organization={IEEE}
}

@inproceedings{rao2020interference,
  title={Interference Characterization in {FMCW} radars},
  author={Rao, Sandeep and Mani, Anil Varghese},
  booktitle={2020 IEEE Radar Conference (RadarConf20)},
  pages={1--6},
  year={2020}
}

@article{hakobyan2019interference,
  title={Interference-aware cognitive radar: A remedy to the automotive interference problem},
  author={Hakobyan, Gor and Armanious, Karim and Yang, Bin},
  journal={IEEE Transactions on Aerospace and Electronic Systems},
  volume={56},
  number={3},
  pages={2326--2339},
  year={2019}
}

@article{gurbuz2019overview,
  title={An overview of cognitive radar: Past, present, and future},
  author={Gurbuz, Sevgi Zubeyde and Griffiths, Hugh D and Charlish, Alexander and Rangaswamy, Muralidhar and Greco, Maria Sabrina and Bell, Kristine},
  journal={IEEE Aerospace and Electronic Systems Magazine},
  volume={34},
  number={12},
  pages={6--18},
  year={2019}
}

@inproceedings{steck2018cognitive,
  title={Cognitive radar principles and application to interference reduction},
  author={Steck, Markus and Neumann, Christoph and Bockmair, Michael},
  booktitle={2018 19th International Radar Symposium (IRS)},
  pages={1--10},
  year={2018},
  organization={IEEE}
}

@article{goppelt2010automotive,
  title={Automotive radar--investigation of mutual interference mechanisms},
  author={Goppelt, Markus and Bl{\"o}cher, H-L and Menzel, Wolfgang},
  journal={Advances in Radio Science},
  volume={8},
  pages={55--60},
  year={2010},
  publisher={Copernicus Publications G{\"o}ttingen, Germany}
}

@article{kim2018peer,
  title={A peer-to-peer interference analysis for automotive chirp sequence radars},
  author={Kim, Geonu and Mun, Jiwoo and Lee, Jungwoo},
  journal={IEEE Transactions on Vehicular Technology},
  volume={67},
  number={9},
  pages={8110--8117},
  year={2018}
}

@inproceedings{fischer2013detection,
  title={Detection of pedestrians in road environments with mutual interference},
  author={Fischer, Christoph and Barjenbruch, Michael and Bloecher, Hans-Ludwig and Menzel, Wolfgang},
  booktitle={2013 14th International Radar Symposium (IRS)},
  volume={2},
  pages={746--751},
  year={2013}
}

@inproceedings{uysal2018mitigation,
  title={Mitigation of automotive radar interference},
  author={Uysal, Faruk and Sanka, Sasanka},
  booktitle={2018 IEEE Radar Conference (RadarConf18)},
  pages={0405--0410},
  year={2018}
}

@inproceedings{yildirim2019impact,
  title={Impact of phase noise on mutual interference of {FMCW} and {PMCW} automotive radars},
  author={Yildirim, Hasan Can and Bauduin, Marc and Bourdoux, Andr{\'e} and Horlin, Fran{\c{c}}ois},
  booktitle={2019 16th European Radar Conference (EuRAD)},
  pages={181--184},
  year={2019},
  organization={IEEE}
}

@article{norouzian2021phenomenology,
  title={Phenomenology of automotive radar interference},
  author={Norouzian, Fatemeh and Pirkani, Anum and Hoare, Edward and Cherniakov, Mikhail and Gashinova, Marina},
  journal={IET Radar, Sonar \& Navigation},
  volume={15},
  number={9},
  pages={1045--1060},
  year={2021},
  publisher={Wiley Online Library}
}

@techreport{buller2018radar,
  title={Radar congestion study},
  author={Buller, William and Wilson, Brian and Garbarino, Joseph and Kelly, Jack and Thelen, Brian and Belzowski, Bruce M and others},
  year={2018},
  institution={United States. Department of Transportation. National Highway Traffic Safety}
}

@article{sae2014taxonomy,
  title={Taxonomy and definitions for terms related to on-road motor vehicle automated driving systems},
  author={SAE On-Road Automated Vehicle Standards Committee and others},
  journal={SAE Standard J},
  volume={3016},
  pages={1},
  year={2014}
}

@article{bengler2014three,
  title={Three decades of driver assistance systems: Review and future perspectives},
  author={Bengler, Klaus and Dietmayer, Klaus and Farber, Berthold and Maurer, Markus and Stiller, Christoph and Winner, Hermann},
  journal={IEEE Intelligent transportation systems magazine},
  volume={6},
  number={4},
  pages={6--22},
  year={2014},
  publisher={IEEE}
}

@inproceedings{zhao2019recent,
  title={Recent development of automotive {LiDAR} technology, industry and trends},
  author={Zhao, Fuquan and Jiang, Hao and Liu, Zongwei},
  booktitle={Eleventh International Conference on Digital Image Processing (ICDIP 2019)},
  volume={11179},
  pages={1132--1139},
  year={2019},
  organization={SPIE}
}

@article{NAAEB,
  title={North {America} Publishes Report on recent Automaker Automatic Emergency Braking Commitment},
  year={2016},
  journal={JATO}
}

@article{EUAEB,
  title={Parliament approves {EU} rules requiring life-saving technologies in vehicles | News | {European} Parliament},
  year={2019},
  journal={Europarl.europa.eu}
}

@article{JapanAEB,
  title={{AEB} to be Required on New Cars in {Japan}},
  year={2019},
  journal={The Brake Report}
}

@article{carullo2001ultrasonic,
  title={An ultrasonic sensor for distance measurement in automotive applications},
  author={Carullo, Alessio and Parvis, Marco and others},
  journal={IEEE Sensors journal},
  volume={1},
  number={2},
  pages={143},
  year={2001}
}

@article{wang2023performance,
  title={Performance analysis of uncoordinated interference mitigation for automotive radar},
  author={Wang, Yi and Zhang, Qixun and Wei, Zhiqing and Kui, Liping and Liu, Fan and Feng, Zhiyong},
  journal={IEEE Transactions on Vehicular Technology},
  volume={72},
  number={4},
  pages={4222--4235},
  year={2023}
}

@inproceedings{mazher2024automotive,
  title={Automotive Radar Interference Characterization: {FMCW} or {PMCW}?},
  author={Mazher, Khurram Usman and Graff, Andrew and Gonz{\'a}lez-Prelcic, Nuria and Heath, Robert W},
  booktitle={ICASSP 2024-2024 IEEE International Conference on Acoustics, Speech and Signal Processing (ICASSP)},
  pages={13406--13410},
  year={2024}
}

@book{richards2005fundamentals,
  title={Fundamentals of radar signal processing},
  author={Richards, Mark A and others},
  volume={1},
  year={2005},
  publisher={Mcgraw-hill New York}
}

@book{skolnik,
  title={Radar Handbook, Third Edition},
  author={Skolnik, Merrill},
  year={2008},
  publisher={Mcgraw-hill New York}
}

@article{deng2013interference,
  title={Interference mitigation processing for spectrum-sharing between radar and wireless communications systems},
  author={Deng, Hai and Himed, Braham},
  journal={IEEE Transactions on Aerospace and Electronic Systems},
  volume={49},
  number={3},
  pages={1911--1919},
  year={2013},
  publisher={IEEE}
}

@article{liu2020decentralized,
  title={Decentralized automotive radar spectrum allocation to avoid mutual interference using reinforcement learning},
  author={Liu, Pengfei and Liu, Yimin and Huang, Tianyao and Lu, Yuxiang and Wang, Xiqin},
  journal={IEEE Transactions on Aerospace and Electronic Systems},
  volume={57},
  number={1},
  pages={190--205},
  year={2020},
  publisher={IEEE}
}

@inproceedings{solodky2021clean,
  title={{CLEAN receiver for CDMA MIMO radar}},
  author={Solodky, Gaston and Longman, Oren and Eljarat, Ishai and Bilik, Igal},
  booktitle={29th European Signal Processing Conference (EUSIPCO)},
  pages={1760--1764},
  year={2021}}
